\documentclass[twocolumn,amsmath,amssymb,aps,]{revtex4-1}
\usepackage{bm}
\usepackage[colorlinks=true,urlcolor=blue,linkcolor=blue,citecolor=blue]{hyperref}
\usepackage{color}
\usepackage{graphics}
\usepackage{graphicx}
\usepackage{epsfig}
\usepackage{amssymb}
\usepackage{amsmath}
\usepackage{hyperref}
\usepackage{physics}
\usepackage[mathscr]{euscript}
\usepackage{ bbold }
\usepackage{hyperref,braket,graphicx,xcolor,amsmath,amssymb}
\newcommand{\dif}[1]{\ensuremath{\mathrm{d}{#1}}}

\usepackage{array}                 
\usepackage{amsfonts}           

\begin{document}

\title{A Coupled-Oscillators Model to Analyze the Interaction between a Quartz Resonator and Trapped Ions}
\author{E.~Altozano$^1$}
\author{J.~Berrocal$^{1}$} 
\author{S.~Lohse$^{2,3,4}$}
\author{F.~Dom\'inguez$^1$}
\author{M.~Block$^{2,3,4}$}
\author{J.J.~Garc\'ia-Ripoll$^5$}
\author{D.~Rodr\'iguez$^{1,6}$}\email[]{danielrodriguez@ugr.es}
\affiliation{
$^1$Departamento de F\'isica At\'omica, Molecular y Nuclear, Universidad de Granada, 18071 Granada, Spain \\
$^2$Department Chemie - Standort TRIGA, Johannes Gutenberg-Universit\"at Mainz, D-55099, Mainz, Germany \\
$^3$GSI Helmholtzzentrum f\"ur Schwerionenforschung GmbH, D-64291, Darmstadt, Germany \\
$^4$Helmholtz-Institut Mainz, D-55099, Mainz, Germany \\
$^5$Instituto de F\'isica Fundamental, Consejo Superior de  Investigaciones Cient\'ificas, Spain\\
$^6$Centro de Investigaci\'on en Tecnolog\'ias de la Informaci\'on y las Comunicaciones, Universidad de Granada, 18071 Granada, Spain
}

\date{\today}

\begin{abstract}
The novel application of a piezoelectric quartz resonator for the detection of trapped ions has developed in the observation of the quartz-ions interaction under non-equilibrium conditions, opening new perspectives for high-sensitive motional frequency measurements of radioactive particles. Energized quartz crystals have (long) constant-decay times in the order of milliseconds, permitting the coherent detection of charged particles within short times. In this publication we develop in detail a model governing the interaction between trapped $^{40}$Ca$^+$ ions and a quartz resonator connected to a low-noise amplifier. We apply this model to experimental data and extract relevant information like the coupling constant $g=2\pi \times 1.449(2)$~Hz and the ions' modified-cyclotron frequency in our 7-tesla Penning trap. The study on the latter is specially important for the use of this resonator in precision Penning-trap mass spectrometry. The improvement in sensitivity can be accomplished by increasing the coupling constant through the quality factor of the resonator. This can develop in the use of the hybrid quartz-ion system for other applications.

\end{abstract}


\maketitle

\section{Introduction}

Electronic detection of trapped charged particles in Penning traps is used in a variety of experiments in fundamental Physics  \cite{Myer2019,Borc2022} as well as applications in Chemistry and Biology \cite{Hend2015}. Most of these experiments pertain to mass spectrometry aiming at reaching the highest precision when only a single ion \cite{Risc2020} or an ion pair \cite{Rain2004} is in the Penning trap, or when the highest precision is not required, aiming at identifying molecular ions using tens or more ions from the same species without frequency-selective amplification. The increase in sensitivity to reach single-ion detection relies mainly on the use of high-$Q$ resonators, which until very recently were made only of superconducting solenoids \cite{Weis1988,Ulme2009}. Recent developments and Penning trap experiments have demonstrated that quartz crystals can serve as resonators  covering, depending on the cut along the crystallographic axis of the crystal, different frequency regimes, from MHz (AT-cut) \cite{Lohs2019} to a few hundred of kHz (SL-cut) \cite{Lohs2020}. The latter work includes the first proof of a cyclotron-frequency ratio measurement of two ion species. Besides, quartz resonators offer a feature not present in superconducting coils, arising from the relatively long decay-constant of a few milliseconds when the quartz is energized by a resonant radiofrequency field. This has resulted in the observation of the coherent interaction between the resonator and trapped ions under non-equilibrium conditions, without the need for long observation times. This has been presented in a recent work \cite{Berr2021} opening prospects for non-destructive detection of exotic nuclei at existing facilities \cite{Bloc2010,Mina2012} or under construction \cite{Rodr2010}. Two models were introduced in Ref.~\cite{Berr2021}, one based on the ions' equivalent circuit \cite{Wine1975}, which is the one applied in precision Penning-trap mass spectrometry \cite{Stur2019}, and the other based on the coupling between the quartz and ions considering each system as a harmonic oscillator. In this manuscript we have fully developed the coupled-oscillators model using the experimental data presented in Ref.~\cite{Berr2021}. We have stablished an analysis procedure which allows us extracting precise motional frequency values from the quartz-ions interaction at different times after the crystal has been energized. We extract important parameters like the coupling constant, subject of  further experimental improvements, and the ions' reduced-cyclotron frequency. We obtain more accurate values compared to the previous determination \cite{Berr2021}, which allows our collaboration to move further in the way to use quartz crystals for precision Penning-trap mass spectrometry on a single ion, a field exploited so far with stable ions, under the domain of superconducting-solenoid based detection setups \cite{Myer2019,Borc2022,Risc2020}. In this respect, single-ion sensitivity, in experiments with quartz crystals, must be reached. From our analysis, this requires an increase of the coupling constant, which depends on trap parameters and on the quality factor of the resonator.

Due to the range of frequencies available, another application in the field of Penning-trap mass spectrometry will be the coupling of ions in physically separated traps to allow using fluorescence photons from a laser-cooled $^{40}$Ca$^+$ (sensor) ion to weight heavy nuclei \cite{Rodr2012}, extending what has been done for the cooling of protons via a cloud of laser-cooled $^{9}$Be$^+$ ions connecting the two traps to an LC circuit \cite{Bohm2021}. Such coupling, and the unique feature of operation under non-equilibrium, might extend the use of quartz crystals and the model presented here beyond mass spectrometry with further exploitation of the hybrid system quartz-ion \cite{Kotl2017}. 

\section{Penning trap and induced image current}

In a Penning trap an ion with mass $m_{\hbox{\scriptsize{ion}}}$ and charge $q$ is confined by the superposition of an electrostatic quadrupole field with a strong homogeneous magnetic field \cite{Brow1986}. The latter defines the revolution-symmetry axis ($\Vec{B}=B\,\vec{e_z}$) of the device made in the simplest configuration by a ring electrode and two endcaps. The configuration of the electrodes is such that the electrostatic potential inside the trap volume, when a voltage $U_0$ is applied between the ring and endcap electrodes, is given by
\begin{equation}
    V\left(x,\,y,\,z\right)=\frac{U_0}{4 d^2}\left(-x^2-y^2+2z^2\right),
\end{equation}
where $d=\frac{1}{2}\sqrt{2 z_0^2+r_0^2}$ and $z_0$ and $r_0$ are the distances from the trap center to the endcap and ring electrode, respectively. The ion motion due to the Lorentz force acting on it, can be depicted as the superposition of three eigenmotions, one in the axial direction with a characteristic frequency
\begin{equation}
\omega_z=\sqrt{\frac{q_{\hbox{\scriptsize{ion}}}U_0}{m_{\hbox{\scriptsize{ion}}}d^2}},
\end{equation}
and two in the radial plane with characteristic frequencies
\begin{equation}
    \omega_{\pm}=\frac{1}{2}\left(\omega_c\pm\sqrt{\omega_c^2-2\omega_z^2}\right)
    \label{eqn:omega_+_-}
\end{equation}
where
\begin{equation}
\omega_c=\frac{q_{\hbox{\scriptsize{ion}}}B}{m_{\hbox{\scriptsize{ion}}}}
\end{equation}
is the cyclotron frequency of the ion. The subscripts + and - in Eq.~(\ref{eqn:omega_+_-}) represent cyclotron motion with the modified frequency and magnetron motion respectively. The quantum Hamiltonian is presented in Appendix~\ref{quantum_description}.

\begin{figure}[t]
\includegraphics[scale=0.7]{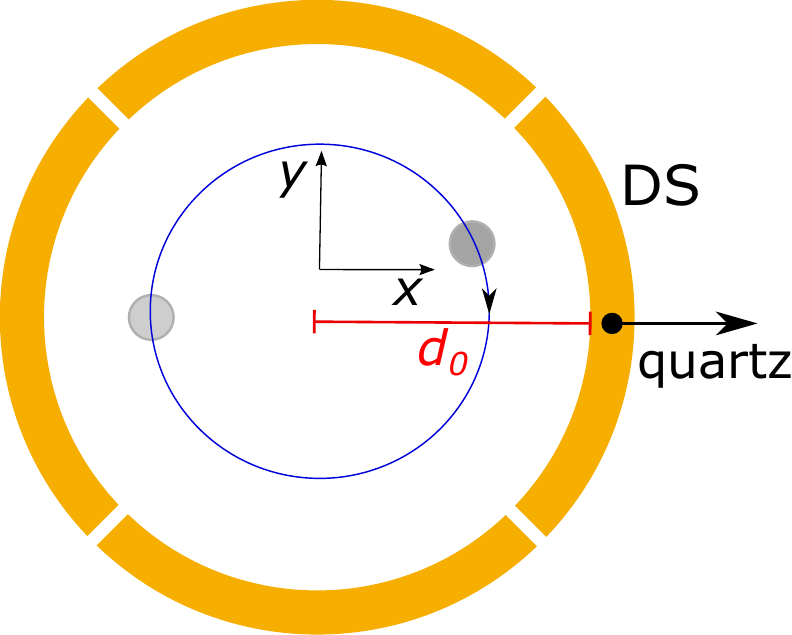}
\vspace{0mm}
\caption{Transverse cut of a four-fold ring segmented electrode in a Penning trap with cylindrical symmetry. The ions' trajectory in the radial plane, assuming only modified-cyclotron motion is depicted by the blue-solid line. The light and dark-grey solid circles represent the trapped ions when they are far or close, respectively, to the detection segment (DS) where the current they induce is picked up. DS is connected to the quartz resonator followed by an amplifier \cite{Lohs2019}.}
\label{fig:ind_charge}
\end{figure}

Any of the characteristic frequencies can be measured if an electrode is used as pick-up current detector, since this current is modulated with the ion motion. Our experiments are centered in the detection of $\omega _+$, and thus the radial motion in the $x$-$y$ plane has to be considered. In first-oder approximation of the radial motion, the charge induced on the detection segment (DS in Fig.~\ref{fig:ind_charge}) by a trapped ion only depends on $x$ (following the nomenclature in Fig.~\ref{fig:ind_charge}) as
\begin{equation}
    Q_{\hbox{\scriptsize{ion}}}(x,y)\approx Q_{\hbox{\scriptsize{ion}}}(x,\,0)\approx-\frac{q_{\hbox{\scriptsize{ion}}}\alpha }{2}\left(1+\frac{x}{d_0}\right),
    \label{eqn:x-axis}
\end{equation}
where $\alpha $ is a geometrical factor that accounts for the deviation of the geometry of the trap electrodes from the ideal case of two parallel plates, and $d_0$ is depicted in Fig.~\ref{fig:ind_charge}. The induced image current is given by
\begin{equation}
    I_{\hbox{\scriptsize{ion}}}(p_x,\,p_y)\approx I_{\hbox{\scriptsize{ion}}} (p_x,0) \approx-\frac{q_{\hbox{\scriptsize{ion}}}\alpha}{2d_0 m_{\hbox{\scriptsize{ion}}}}p_x.
    \label{eqn:induced_charge}
\end{equation}

\section{The quartz-ion coupled-oscillator model}\label{sec:model}

\begin{figure}[t]
\includegraphics[scale=0.7]{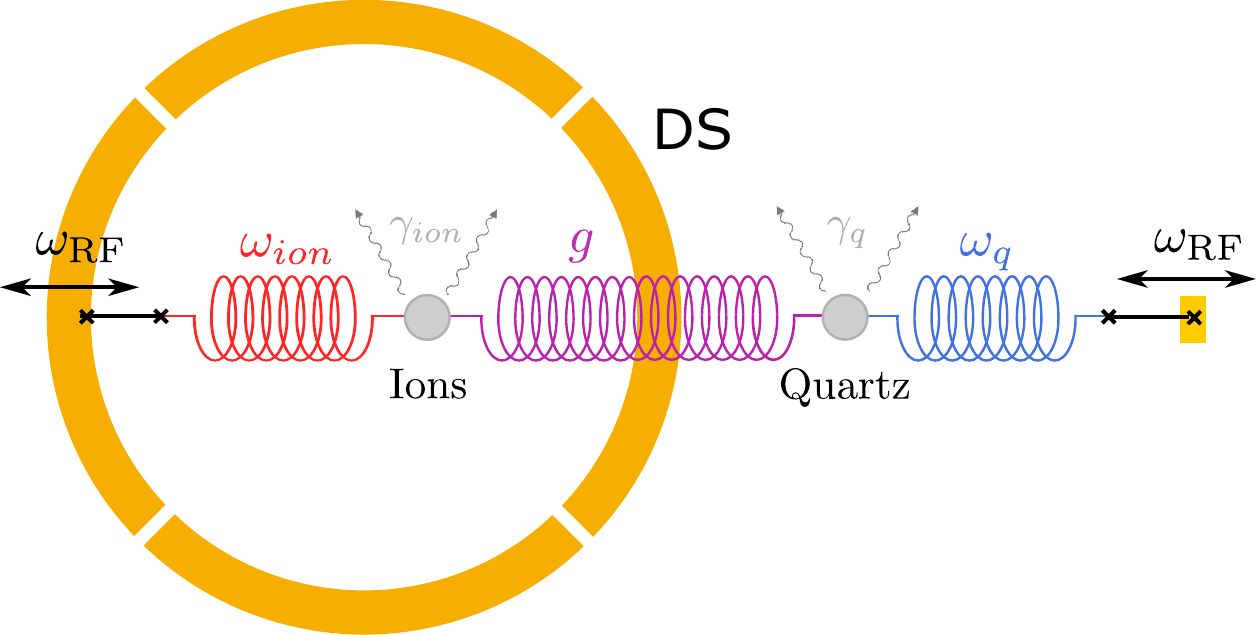}
\vspace{-5mm}
\caption{Schematic view of the quartz-ion coupled-oscillators model.}
\label{fig:model}
\end{figure}

The full system is depicted in Fig.~\ref{fig:model} and the associated Hamiltonian comprises the following contributions
\begin{equation}
 H_{\hbox{\scriptsize{total}}}= H_{\hbox{\scriptsize{ion}}} + H_{\hbox{\scriptsize{q}}} + H_{\hbox{\scriptsize{int}}} + H_{\hbox{\scriptsize{RF}}} \label{eq:total_hamiltonian}
 \end{equation}
where $H_{\hbox{\scriptsize{q}}}$ describes the dynamics of the quartz crystal, $H_{\hbox{\scriptsize{ion}}}$ that of the trapped ions, $H_{\hbox{\scriptsize{int}}}$ the interaction between the quartz and the ions, and $H_{\hbox{\scriptsize{RF}}}$ the interaction between both with an external radiofrequency (RF) driving field. All these terms will be described in the following, giving them as a function of the creation and annihilation operators $a^\dagger$ and $a$, for the ions, and $b^\dagger$ and $b$ for the quartz. 

The Hamiltonian to describe the ion motion in a Penning trap is presented in App.~\ref{quantum_description_ion}. Since only the modified-cyclotron motion is in resonance with the quartz, one can write
\begin{equation}
         H_{\hbox{\scriptsize{ion}}}\approx\hbar\omega_{+}\left(a_{+}^{\dagger}a_{+}+\frac{1}{2}\right)\equiv \hbar\omega_{\hbox{\scriptsize{ion}}}\left(a^{\dagger}a+\frac{1}{2}\right)  
           \label{eq:H_ions_rad_simp}
\end{equation}
with $\langle a \rangle \equiv \abs{\langle a\rangle }e^{i\theta _{a}}$.

The Hamiltonian describing the quartz is given by

\begin{equation}
        H_{\hbox{\scriptsize{q}}}=\left(\frac{k}{C_{\hbox{\scriptsize{q}}}}\right)^{2}\left(\frac{m_{\hbox{\scriptsize{q}}}}{2}\hat{I}^{2}+\frac{m_q\omega_{\hbox{\scriptsize{q}}}^{2}}{2}\hat{Q}^{2}\right)=\hbar\omega_{\hbox{\scriptsize{q}}}\left(b^\dagger b+\frac{1}{2}\right),
    \label{eq:H_q}
\end{equation}
with 
\begin{equation}
               b=\sqrt{\frac{m_{\hbox{\scriptsize{q}}}\omega_{\hbox{\scriptsize{q}}}}{2\hbar}}\frac{k}{C_{\hbox{\scriptsize{q}}}}\left[\hat{Q}+\frac{i}{\omega_{\hbox{\scriptsize{q}}}}\hat{I}\right],
    \label{eq:b_operator}
\end{equation}
and $\langle b\rangle \equiv |\langle b\rangle |e^{i\theta _{b}}$ its expected value. The operators $\hat Q$ and $\hat I$ related to charge and current are defined in App.~\ref{quantum_description_quartz}. $C_{\hbox{\scriptsize{q}}}$ is the capacitance of the quartz resonator,  $m_q$ the mass of a quartz's oscillation mode with resonant frequency $\omega_{\hbox{\scriptsize{q}}}$, and $k$ is a proportionally constant which relates the expansion of the piezoelectric crystal with the applied voltage.

Following Appendix~\ref{quantum_description_int}, and considering that the magnetron motion is not in resonance, the Hamiltonian to describe the interaction can be written in the rotating wave approximation as
\begin{equation}
    \begin{split}
        H_{\hbox{\scriptsize{int}}}&=\left(\frac{k}{C_{\hbox{\scriptsize{q}}}}\right)^{2}\left(-\frac{Nq_{\hbox{\scriptsize{ion}}}\alpha}{4d_{0}m_{\hbox{\scriptsize{ion}}}m_{\hbox{\scriptsize{q}}}}\hat{I}\hat{p}_{x}-\frac{Nq_{\hbox{\scriptsize{ion}}}\alpha m_{\hbox{\scriptsize{q}}}\omega_{\hbox{\scriptsize{q}}}^{2}}{4d_{0}}\hat{Q}\hat{x}\right)\\
        &\approx \hbar g a^\dagger b + \hbar  g^* b^\dagger a, \label{eq:H_int}
    \end{split}
\end{equation}
where $N$ is the number of trapped ions and $g/2\pi$ represents the coupling constant in Hz. Taking into account that $\omega_-\ll \omega_+$ and $\omega_{\hbox{\scriptsize{q}}}\approx\omega_+$, the coupling constant can be written as
\begin{equation}
 \begin{split}
     g=i\frac{k}{C_{\hbox{\scriptsize{q}}}}\frac{ q_{\hbox{\scriptsize{ion}}}\alpha\left(\omega_{\hbox{\scriptsize{q}}}+2\omega_{-}\right)}{8d_{0}}\sqrt{\frac{Nm_{\hbox{\scriptsize{q}}}\omega_{\hbox{\scriptsize{q}}}}{m_{\hbox{\scriptsize{ion}}}\left(\omega_{+}-\omega_{-}\right)}}\\  \approx  i\frac{k}{C_q}\frac{ q_{\hbox{\scriptsize{ion}}}\alpha\omega_{+}}{8d_{0}}\sqrt{\frac{Nm_{\hbox{\scriptsize{q}}}}{m_{\hbox{\scriptsize{ion}}}}}. \,\,\,\,\,\,\,\,\,\,\,\,\,\,\,\,\,\,\,\,\,\,\,\,\,\,\,\,\,\,\,\,\,\,\,\,\,\,\,\,\,\,\,\,\,\,\,&
          \end{split}
     \label{eq:acoplamiento_g}
\end{equation}

The Hamiltonian describing the interaction between the ions and an external RF field reads
\begin{equation}
    \begin{split}
        H_{\hbox{\scriptsize{RF}}}= & \frac{F_\text{ion}}{\sqrt{2}}\cos\left(\omega_{\hbox{\scriptsize{RF}}}t+\phi_{\hbox{\scriptsize{ion}}}\right) (a+a^\dagger) \notag\\
  & + \frac{F_\text{q}}{\sqrt{2}}\cos\left(\omega_{\hbox{\scriptsize{RF}}}t+\phi_{\hbox{\scriptsize{q}}}\right)(b+b^\dagger).\notag
      \end{split}
\end{equation}
where $\omega_{\hbox{\scriptsize{RF}}}$ is the frequency of the field, $F_{\hbox{\scriptsize{ion}}}$  and $F_q$ are the amplitudes of the forces exerted on the ion and on the quartz, respectively, and $\phi_{\hbox{\scriptsize{ion}}}$ and $\phi_{\hbox{\scriptsize{q}}}$ are the phase differences between the forces and operators.

Working in the Heisenberg picture, the operators $a$ and $b$ oscillate with frequencies $e^{-i\omega_\text{ion}t}$ and $e^{-i\omega_\text{q}t}$, respectively. Using the rotating wave approximation 
\begin{align}
  \frac{F_\text{ion}}{\sqrt{2}}\cos\left(\omega_{\hbox{\scriptsize{RF}}}t+\phi_{\hbox{\scriptsize{ion}}}\right) (a+a^\dagger) \sim \\
  \frac{F_\text{ion}}{\sqrt{2}}(e^{i(\omega_{\hbox{\scriptsize{RF}}} t + \phi_{\hbox{\scriptsize{ion}}})}a+\mathrm{H.c.}),
\end{align}
and
\begin{align}
  H_\text{total} = & \hbar \tilde\omega_\text{ion}a^\dagger a + \hbar \tilde\omega_\text{q} b^\dagger b\notag\\
  & + \hbar f_\text{ion}(e^{i\phi_{\hbox{\scriptsize{ion}}}}a+\mathrm{H.c})
    + \hbar f_\text{q}(e^{i\phi_{\hbox{\scriptsize{q}}}}b+\mathrm{H.c.})\\
  &+ \hbar g a^\dagger b + \hbar  g^* b^\dagger a\notag,
\end{align}
with $\tilde\omega_\text{ion}=\omega_\text{ion}-\omega_{\hbox{\scriptsize{RF}}}$ and $\tilde\omega_\text{q}=\omega_\text{q}-\omega_{\hbox{\scriptsize{RF}}}$, and some effective forces $f = F/\hbar\sqrt{2}$.

\section{Power spectral density computation}

The fundamental model for the hybrid ion-quartz-resonator system is a master equation for two coupled oscillators under finite-temperature baths
\begin{equation}
  i\frac{d}{dt}\rho = -\frac{i}{\hbar}[H_{\hbox{\scriptsize{total}}}, \rho] + \mathcal{L}(\rho)
   =\mathcal{L}_\text{tot}(\rho,t)\label{Lindblad},
\end{equation}
with the coherent interaction (Eq.~(\ref{eq:total_hamiltonian})) and a heating term
\begin{align}
  \mathcal{L}\rho =
  & \gamma_q n_q \left(b^\dagger \rho b - \frac{1}{2}b b^\dagger\rho-\frac{1}{2}\rho b b^\dagger\right) \label{eq:linblad}\\
    & + \gamma_n (n_q + 1) \left(b \rho b^\dagger - \frac{1}{2}b^\dagger b\rho-\frac{1}{2}\rho b^\dagger b\right),\notag 
\end{align}
where
\begin{equation}
n_{\hbox{\scriptsize{q}}}=\frac{1}{e^{\hbar \omega_{\hbox{\scriptsize{q}}}/k_BT_{\hbox{\scriptsize{q}}}}-1}\approx \frac{k_BT_{\hbox{\scriptsize{q}}}}{\hbar \omega _{\hbox{\scriptsize{q}}}}
\end{equation} 
is the phonon number in the termal bath of the quartz and $\gamma_{\hbox{\scriptsize{q}}}$ is a dissipative constant to account for the interaction with this environment. $\gamma_{\hbox{\scriptsize{q}}}$ is related to the quality factor $Q$ of the resonator by $\gamma_{\hbox{\scriptsize{q}}}=\omega_{\hbox{\scriptsize{q}}}/Q$. Likewise, $n_{\hbox{\scriptsize{ion}}}$ is the number of phonons in the thermal bath of the ions, $\gamma_{\hbox{\scriptsize{ion}}}$ is the dissipative constant of the ion cloud due to phenomena not described by $H_{\hbox{\scriptsize{total}}}$ like e.g., internal degrees of freedom in the cloud, or interaction with other electrodes or collisions.

Equation~(\ref{eq:linblad}) might be also written as
\begin{align}
  \mathcal{L}\rho =
  & \frac{\gamma_q}{2} n_q \left([b^\dagger,\rho b]+[b^\dagger\rho,b]\right)\\
  & + \frac{\gamma_n}{2} (n_q + 1) \left([b, \rho b^\dagger]+[b\rho,b^\dagger]\right).\notag
\end{align}

The master equation is linear. We can therefore write formally its solution for an initial condition $\rho(t_0)$ as positive map $\rho(t)=\varepsilon[\rho(t_0),t,t_0]$ with some expression $\varepsilon[\cdot]$ to be determined. In absence of driving, with a time-independent Lindblad operator $\mathcal{L}_\text{tot}$, the positive map is independent of the initial condition and only depends on the time interval that we investigate
\begin{equation}
  \rho(t) = e^{t \mathcal{L}_\text{tot} } \rho(0) = \varepsilon_t[\rho(t_0)].
\end{equation}

The experiment drives the ion-oscillator system up to a point. It then measures the voltage on the quartz oscillator, obtaining traces $V_i(t)$ which are Fourier transformed
\begin{equation}
  \tilde{V}_i(\omega;t_0,t_1) = \frac{1}{\sqrt{t_1-t_0}}\int_{t_0}^{t_1}e^{-i\omega t} V_i(t)\dif{t}.
\end{equation}
The power spectrum is approximated by the average of these quantities, as in
\begin{equation}
  S_V(\omega;t_0,t_1,N) :=  \frac{1}{N} \sum_{i=1}^N \left|\tilde{V}_i(\omega;t_0,t_1)\right|^2.
\end{equation}
This can be related to any quantum observable in the combined ion-oscillator system. From the original time traces
\begin{equation}
  S_V(\omega;t_0,t_1,N) = \int_{t_0}^{t_1}\!\!\int_{t_0}^{t_1}\!\!\dif{\tau}_1 \dif{\tau}_2e^{i\omega (\tau_1-\tau_2)}  \sum_{i=1}^N\frac{V_i(\tau_1)V(\tau_2)}{t_d N},
\end{equation}
with $t_d=t_1-t_0$, we assimilate this estimator to a power spectrum estimate over the quantum state, in the limit $N\to\infty$
\begin{equation}
  S_V(\omega;t_0,t_1) := \frac{1}{t_d} \int_{t_0}^{t_1}\!\!\int_{t_0}^{t_1}\!\!\dif{\tau}_1 \dif{\tau}_2e^{i\omega (\tau_1-\tau_2)}  \braket{\hat{V}(\tau_2)\hat{V}(\tau_1)}.
  \label{eq:power-spectrum}
\end{equation}
In the limit of infinite time $t_d$ and stationary systems, this quantity approaches the usual definition of power spectrum \cite{quantum_noise_introduction}
\begin{equation}
  \lim_{t_d\to\infty} S_V(\omega;t_0,t_1) := \int_{-\infty}^\infty e^{-i\omega\tau}\braket{\hat{V}(\tau)\hat{V}(0)}.
\end{equation}

In order to match the experiments and the theory, one needs to compute two-time correlators $\braket{\hat{V}(t_1)\hat{V}(t_2)}$ for the ion-oscillator system. This will be approached in a slightly more general way, and the correlators $\braket{A_n(t)A_m(t')}$ will be computed for a collection of Fock operators that form a suitable basis for all the observables of interest
\begin{equation}
  \mathbf{A} = \begin{pmatrix} a \\ b \\ a^\dagger \\ b^\dagger \end{pmatrix}.
\end{equation}
Using these operators, one can reconstruct the observables of interest
\begin{equation}
  \hat{V} = V_0 \frac{1}{\sqrt{2}}(b + b^\dagger) = \mathbf{v}^{\hbox{\scriptsize{T}}} \cdot \mathbf{A}.
\end{equation}
In these canonical coordinates, we recover the power spectrum
\begin{equation}
  S_V(\omega;t_0,t_1) = \mathrm{Re} \sum_{m,n}v_mv_nS_{mn}(\omega),
\end{equation}
from a slightly more general power spectrum matrix
\begin{equation}
  S_{mn}(\omega;t_0,t_1) :=
  \int_{t_0}^{t_1}\!\!\!\int_{t_0}^{t_1}\!\!\dif{\tau}_1 \dif{\tau}_2 \frac{e^{i\omega (\tau_1-\tau_2)}}{t_d}  \braket{A_m(\tau_2)A_n(\tau_1)}.
\end{equation}

Our way of proceeding involves three steps. The first step is to recover  of the $\mathbf{A}$ operators while the ion and the quartz are driven. This allows us to obtain the initial conditions for the spectroscopy experiment, right at time $t=0$ where the driving is switched off. The second step is to solve  of the correlators during the spectroscopy phase, so that we can gather the statistics of the voltage traces from $t_0$ up to $t_1=t_0+t_d$. The third step is to Fourier transform the resulting correlators to recover Eq.~\eqref{eq:power-spectrum}.

\subsection{Driven ion-oscillator dynamics}

The basis of operators $\mathbf{A}$ almost satisfies the conditions for the quantum regression theorem \cite{breuer}. 
First of all, their expectation values evolve linearly in the basis of operators spanned by $\mathbf{A}$,
\begin{equation}
  \mathrm{tr}\left\{A_n \mathcal{L}_\text{tot}(O)\right\} = M_{nm} \mathrm{tr}(A_m O) + F_n,
  \label{eq:qrt}
\end{equation}
for any operator $O$. This allows writing dynamical equations for the first order moments
\begin{equation}
  \frac{d}{dt}\braket{\mathbf{A}(t)} = -\mathbf{M}\braket{\mathbf{A}(t)} + \mathbf{F}. \label{eq:numerically}
\end{equation}
If one defines the displaced operators
\begin{equation}
  \mathbf{A} := \mathbf{A}^0 + \mathbf{A}_\mathbf{F},\mbox{ with }
  \mathbf{A}_\mathbf{F} := -\mathbf{M}^{-1}\mathbf{F},
\end{equation}
they do satisfy a linear differential equation without source term, which is the requirement for the quantum regression theorem
\begin{equation}
  \frac{d}{dt}\braket{\mathbf{A}^0(t)} = -\mathbf{M}\braket{\mathbf{A}^0(t)}.
\end{equation}

In both equations the evolution is generated by a block-diagonal matrix
\begin{equation}
  \label{eq:qrt-M}
  \mathbf{M} =
  \begin{pmatrix}
    \mathbf{m} & 0 \\ 0 & \mathbf{m}^*
  \end{pmatrix},\mbox{ with }
  \mathbf{m} =
  \begin{pmatrix}
    \frac{\gamma_\text{ion}}{2}+i\tilde\omega_\text{ion}
    & i g^* \\
    i g
    &  \frac{\gamma_\text{q}}{2}+i\tilde\omega_\text{q}
  \end{pmatrix},
\end{equation}
and a force term, which is constant because of the rotating frame 
\begin{equation}
  \mathbf{F} =
  \begin{pmatrix}
    f_\text{ion}e^{-i\phi_{\hbox{\scriptsize{ion}}}} \\ f_\text{q}e^{-i\phi_{\hbox{\scriptsize{q}}}} \\
    f_\text{ion}e^{i\phi_{\hbox{\scriptsize{ion}}}} \\ f_\text{q}e^{i\phi_{\hbox{\scriptsize{q}}}}
  \end{pmatrix}.
\end{equation}

The dynamics of the quartz-ion system can be solved by assuming that the force is switched on at time  $t_s$ and the driving continues until time $t$\begin{align}
  \braket{\mathbf{A}^0(t)}
  &= \mathbf{U}(t-t_s)\braket{\mathbf{A}^0(t_s)}\\
  \braket{\mathbf{A}(t)}&= e^{-\mathbf{M}(t-t_s)}\braket{\mathbf{A}(t_s)} + (\openone - e^{-\mathbf{M}(t-t_s)})\mathbf{A}_\mathbf{F},\notag
\end{align}
using the $4\times 4$ contractive matrix $\mathbf{U}(t) = \exp(-\mathbf{M}t)$. In the limit $t\to+\infty$ this dynamics brings the ion and quartz to a unique stationary solution
\begin{equation}
 \lim_{t\to+\infty} \braket{\mathbf{A}(t)} = - \mathbf{M}^{-1}\mathbf{F} = \mathbf{A}_\mathbf{F}.
\end{equation}

The asymptotic state of the driving $\mathbf{A}_\mathbf{F}$ depends on the inverse of the matrix $\mathbf{M}=\mathbf{M}(\omega_{\hbox{\scriptsize{RF}}})$, which itself depends on the detunings between the drive and the ion/resonator frequencies $\omega_\text{ion}-\omega_{\hbox{\scriptsize{RF}}}$ and $\omega_\text{q}-\omega_{\hbox{\scriptsize{RF}}}$. This produces the usual Lorentzian profile of $\mathbf{A}_\mathbf{F}$ as a function of $\omega$.

In the experiment, the driving is stoped at a time $t=0 $. The dynamics of the ion and the quartz naturally revert to the dissipation induced friction, and the displacement of the Fock operators (and the average values of voltage and intensity) naturally revert back to zero
\begin{equation}
  \braket{\mathbf{A}(t)} = \mathbf{U}(t) \mathbf{A}_\mathbf{F} \to 0,\mbox{ as }
  t\to\infty.
    \label{eq:displacement-decay}
\end{equation}

The study of two-time correlators needs of higher-order moments of the ion-quartz system. With similar effort, a linear differential equation for the same-time correlators $\braket{A_m(t)A_n(t)}$ can be computed
\begin{align}
  \label{eq:correlator-dynamics}
  \frac{d}{dt}\braket{A_nA_m} =
  &-M_{mr}\braket{A_rA_n} + F_m\braket{A_n} + \\
  &-M_{ns}\braket{A_mA_s} + \braket{A_m}F_n + C_{mn}\notag,
\end{align}
with a real matrix $C_{nm}$ [cf App.~\ref{correlator}]. This equation has a unique stationary state satisfying
\begin{equation}
M_{mr}\braket{A^0_rA^0_n} + M_{ns}\braket{A^0_mA^0_s} +C_{mn}=0.\label{eq:static-correlation}
\end{equation}
with the displaced operators $\mathbf{A}^0 = \mathbf{A} + \mathbf{M}^{-1}\mathbf{F}$.

The solution to Eq.~\eqref{eq:static-correlation} is the correlation matrix of a thermal state, dictated by the $n_\text{ion}$ and $n_\text{q}$ occupations that define $C_{mn}$. This means that the stationary solution during the driving can be written as two contributions: a thermal state, and a displacement induced by the driving
\begin{equation}
  \braket{A_nA_m} = \braket{A_nA_m}_\text{th} +
   \braket{A_F}_m  \braket{A_F}_n.
  \label{eq:driven-correlator}
\end{equation}

After stopping the driving at $t=0$, the dynamics of the same-time correlator follows a similar route as the first order moment, with a constant term, given by the thermal fluctuations, and a product of two displacements
\begin{equation}
  \label{eq:same-time-correlator}
  \braket{A_n(t)A_m(t)} = \braket{A_nA_m}_\text{th} + \braket{A_m(t)}\braket{A_n(t)},
\end{equation}
which attenuate as described above in Eq.~\eqref{eq:displacement-decay}.

\subsection{Free dynamics of two-time correlators}

Once the driving stops, the $\mathbf{A}$ operators satisfy a differential equation without source term that can be solved linearly
\begin{equation}
  \braket{\mathbf{A}(t)} = \mathbf{U}(t-t_0) \braket{\mathbf{A}(t_0)},\;t,t_0>0.
\end{equation}
This allows using the quantum regression theorem to derive two-time correlators at all possible orderings of time $t,\tau>0$ after switching off the driving
\begin{align}
  \braket{A_m(t+\tau) A_n(t)} &= U_{mr}(\tau)\braket{A_r(t)A_n(t)},\\
  \braket{A_m(t) A_n(t+\tau)} &= U_{nr}(\tau)\braket{A_m(t)A_r(t)}.
\end{align}

Given the simple structure of the same-time correlator~\eqref{eq:same-time-correlator}, we can derive an explicit formula for the two-time correlator we use in the power spectrum. This formula has a simple contribution given by the global displacement, and a slightly more complex one given by the thermal fluctuations
\begin{align}
  \label{eq:two-time-correlator}
  &\braket{A_m(t_2)A_n(t_1)}
    = \braket{A_m(t_2)}\braket{A_n(t_1)}+\\
  &\quad\quad+
    \left\{\begin{array}{ll}
             U_{mr}(t_2-t_1)\braket{A_rA_n}_\text{th} & \mbox{if } t_2 > t_1,\\
             U_{nr}(t_1-t_2)\braket{A_mA_r}_\text{th} & \mbox{else}.\\
           \end{array}
    \right.\notag
\end{align}

\subsection{Power spectrum}

From Eq.~(\ref{eq:two-time-correlator}), the power spectrum can be written as the sum of two terms, a thermal noise plus the coherent signal that results from the damped oscillations in both the quartz resonator and the ions. One can also add a frequency independent term arising from the electronic noise of the circuit, so that the total power can be written as
\begin{equation}
  S(\omega) = S^{\hbox{\scriptsize{coh}}}(\omega)+S^{\text{th}}(\omega) + S^{\text{noise}}. \label{eq:fitting_function_general}
  \end{equation}
The coherent part of the power spectrum may be derived from the dynamics of the first order moments
\begin{align}
  &S^\text{coh}(\omega;t_1,t_0)_{mn} = \tilde{A}_m(\omega;t_1,t_0) \tilde{A}_n(-\omega;t_1,t_0),
\end{align}
as the Fourier transform of the first order moments
\begin{equation}
  \tilde{A}_m(\omega;t_1,t_0) = \frac{1}{\sqrt{t_d}}\int_{t_0}^{t_1}
  e^{-i\omega t} \braket{A_m(t)}\dif{t}.
\end{equation}
Given the exact dynamics for the first order moments~\eqref{eq:displacement-decay}, these integrals can be computed [cf. App.~\ref{sec:propagator}], obtaining
\begin{align}
  \mathbf{\tilde{A}}&=(-i\omega - \mathbf{M})^{-1} \left(e^{-i\omega t_d}\braket{\mathbf{A}(t_1)} - \braket{\mathbf{A}(t_0)}\right).\notag
\end{align}

The voltage power spectrum is the sum of four terms in the $S_{mn}$ matrix
\begin{align}
  &S^\text{coh}_V(\omega;t_0+t_d,t_0) =\\
  &\quad\notag\frac{1}{2t_d}\left[\tilde{A}_2(\omega) + \tilde{A}_4(\omega)\right]\left[\tilde{A}_2(-\omega) + \tilde{A}_4(-\omega)\right].
\end{align}
This can be simplified. Due to the organization of the $\mathbf{\tilde{A}}$ vector, $\tilde{A}_4(\omega) = \tilde{A}_2(-\omega)^*$. In addition, for $\omega>0$, only the lower-right sector of the inverse of $(-i\omega-\mathbf{M})$ survives, hence $\tilde{A}_4(-\omega) = \tilde{A}_2(\omega)^*\simeq 0$ whenever $\omega > 0$. Thus
\begin{equation}
  S^\text{coh}_V(\omega;t_0+t_d,t_0) = \frac{V_0^2}{2t_d}|\tilde{A}_4(\omega)|^2.
\end{equation}
where
\begin{align} \label{eq:coherent_power}
  &\tilde{A}_4(\omega)=  -F(\omega)^*ig\left(e^{-i\omega t_d}\braket{a^\dagger(t_1)} - \braket{a^\dagger(t_0)}\right) \\
  & -F(\omega)^*\left[i(\omega-\omega_\text{ion})+\frac{\gamma_\text{ion}}{2}\right] 
   \left(e^{-i\omega t_d}\braket{b^\dagger(t_1)} - \braket{b^\dagger(t_0)}\right),\notag 
\end{align}
with the envelope
\begin{equation}
  F(\omega)^* =\frac{1}{\left[i(\omega-\omega_\text{ion})+\frac{\gamma_\text{ion}}{2}\right]
  \left[i(\omega-\omega_\text{q})+\frac{\gamma_\text{q}}{2}\right]
  + |g|^2}.
\label{eq:envelope}
\end{equation}

The thermal component of the power spectrum can be written as the real part of just one matrix integral [cf. App.~\ref{sec:power-spectrum-matrix}] 
\begin{align}
  \label{eq:thermal-spectrum}
  S^\text{th}_V(\omega) &= 2\mathrm{Re}\,\sum_{mn} v_m v_n S_{mn}^+(\omega),\mbox{ with}\\
  S_{mn}^+(\omega) &=\frac{1}{t_d} \int_{t_0}^{t_1}\!\!\dif{t}\int_{0}^{t_d-t}\!\!\dif{\tau} e^{-i\omega \tau}
    U_{mr}(\tau)\braket{A_rA_n}_\text{th}.
\end{align}
Since the expectation value inside the integral is constant, one only needs to compute the matrix
\begin{align}
  \mathbf{Q}(\omega) =
  &\frac{1}{t_d}\int_{0}^{t_d}\dif{t}\int_{0}^{t_d-t}\dif{\tau} e^{(-i\omega-\mathbf{M}) \tau}\\
  =&\frac{1}{t_d}(-i\omega-\mathbf{M})^{-2} \left(e^{(-i\omega-\mathbf{M})t_d}-\openone\right) + \\
  &+(-i\omega-\mathbf{M})^{-1}\frac{1}{t_d}\times t_d\notag\\
  \simeq& (-i\omega - \mathbf{M})^{-1}\mbox{ as } t_d\to+\infty.
\end{align}

Given this simplification in the limit of long integration times, the thermal power spectrum is
\begin{equation}
  \mathbf{S}^+(\omega) = 2\left[(-i\omega - \mathbf{M})^{-1}\mathbf{T}\right]_{mr},
\end{equation}
with the matrix of static correlations from Eq.~\eqref{eq:thermal}. Since $\mathbf{T}$ is only nonzero in two sectors, and since we know the explicit expressions for the inverse of $\mathbf{M}$, one can write
\begin{align}
  \mathbf{S}^+(\omega) \simeq
  \begin{pmatrix}
    0 & F(-\omega)\mathbf{G}(-\omega) \mathbf{t}_{-\dagger} \\
    F(\omega)^*\mathbf{G}(\omega)^* \mathbf{t}_{-\dagger} & 0
  \end{pmatrix}.
\end{align}

As before, $F(-\omega)\simeq 0$ for positive values of $\omega$. Out this mostly zero matrix, only one element contributes to the autocorrelation of the potential
\begin{equation}
  S^\text{th}_V (\omega) = \frac{V_0^2}{2}\times 2 \mathrm{Re}\left[S^+_{b^\dagger b}(\omega)\right]
\end{equation}
Introducing the explicit formula for $\mathbf{G}(\omega)$, one obtains
\begin{align} \label{eq:thermal_power}
  S^+_{bb^\dagger} (\omega) &= -F(\omega)^*ig\braket{a^\dagger b}_\text{th} \\
  &- F(\omega)^*\left[\frac{\gamma_\text{ion}}{2} + i (\omega - \omega_\text{ion})\right]\braket{b^\dagger b}_\text{th},\notag 
\end{align} 
where $\braket{a^\dagger b}_\text{th}$ and $\braket{b^\dagger b}_\text{th}$ are given in App.~\ref{thermal_state_appendix}.

\section{Results and discussions}

\begin{figure}[t]
\includegraphics[scale=0.33]{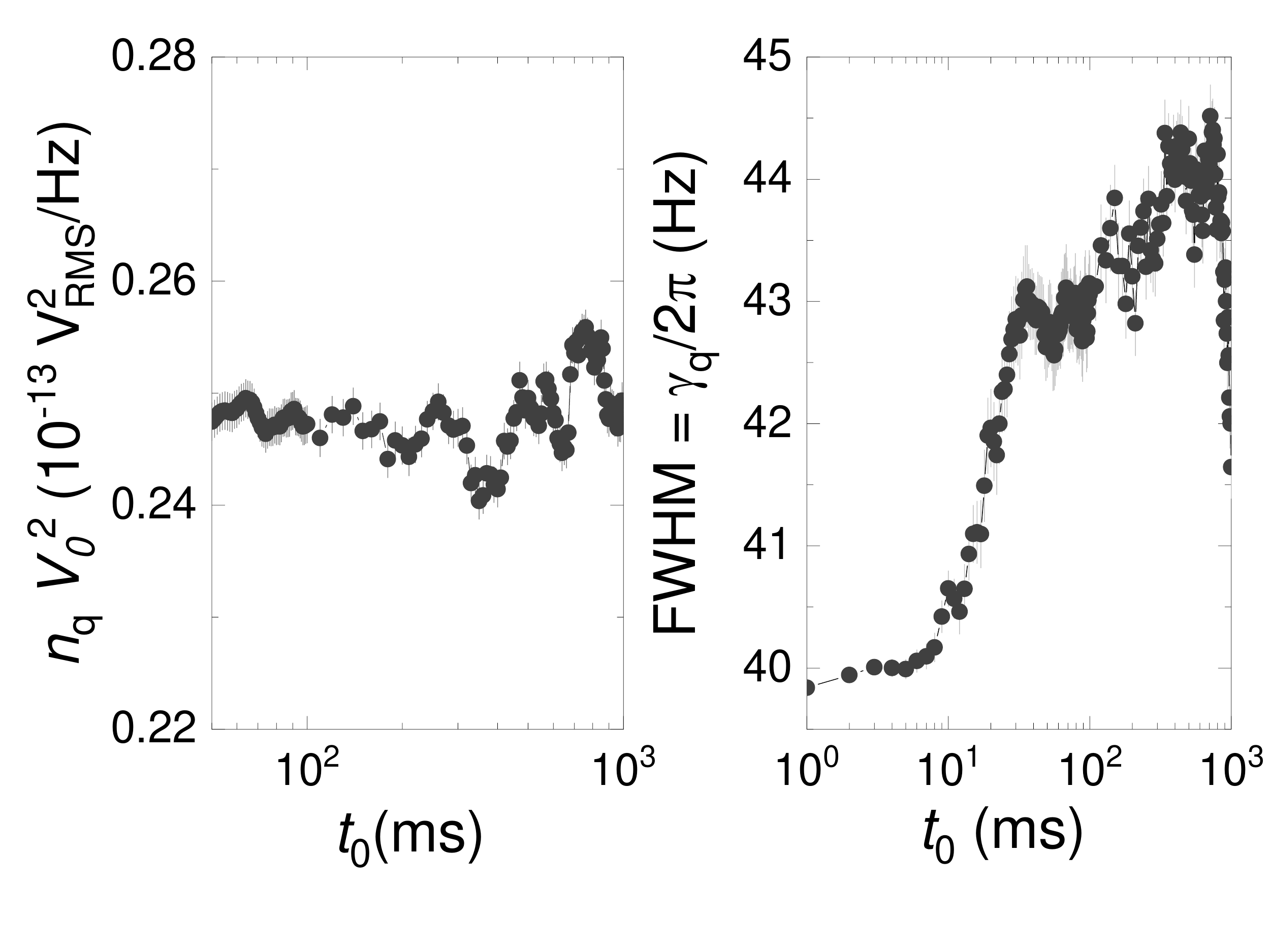}
\vspace{-8mm}
\caption{Left panel: $n_{\hbox{\scriptsize{q}}}$ as a function of the time $t_0$. Right panel: $\gamma _{\hbox{\scriptsize{q}}}$ as a function of the time $t_0$.}
\label{fig:fit_Nq}
\end{figure}

\begin{figure}[t]
\includegraphics[scale=0.35]{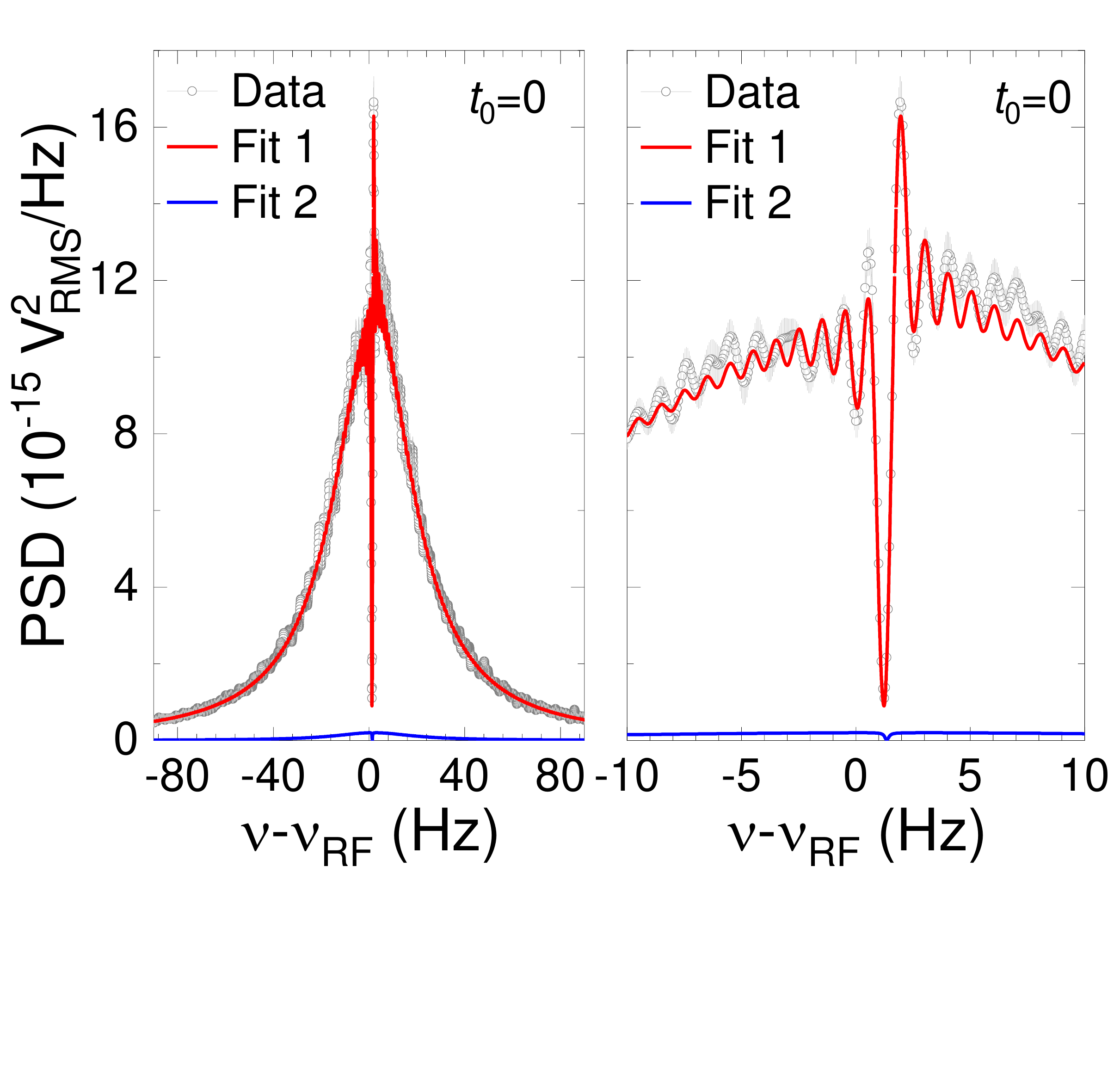}
\vspace{-23mm}
\caption{PSD signal as a function of the frequency for \linebreak $t_0=0$~ms. A zoomed plot is shown in the right panel. The red-solid line in both panels (Fit 1) is the $S(\omega)$ fit. The blue-solid line (Fit 2) is the $S^{\text{th}}(\omega) + S^{\text{noise}}$ fit. $\chi ^2 _{\nu } = 1.18$. The data points and standard deviations (1~$\sigma $) are the results from 20 measurements.}
\label{fig:fit1}
\end{figure}

\begin{figure}[t]
\includegraphics[scale=0.4]{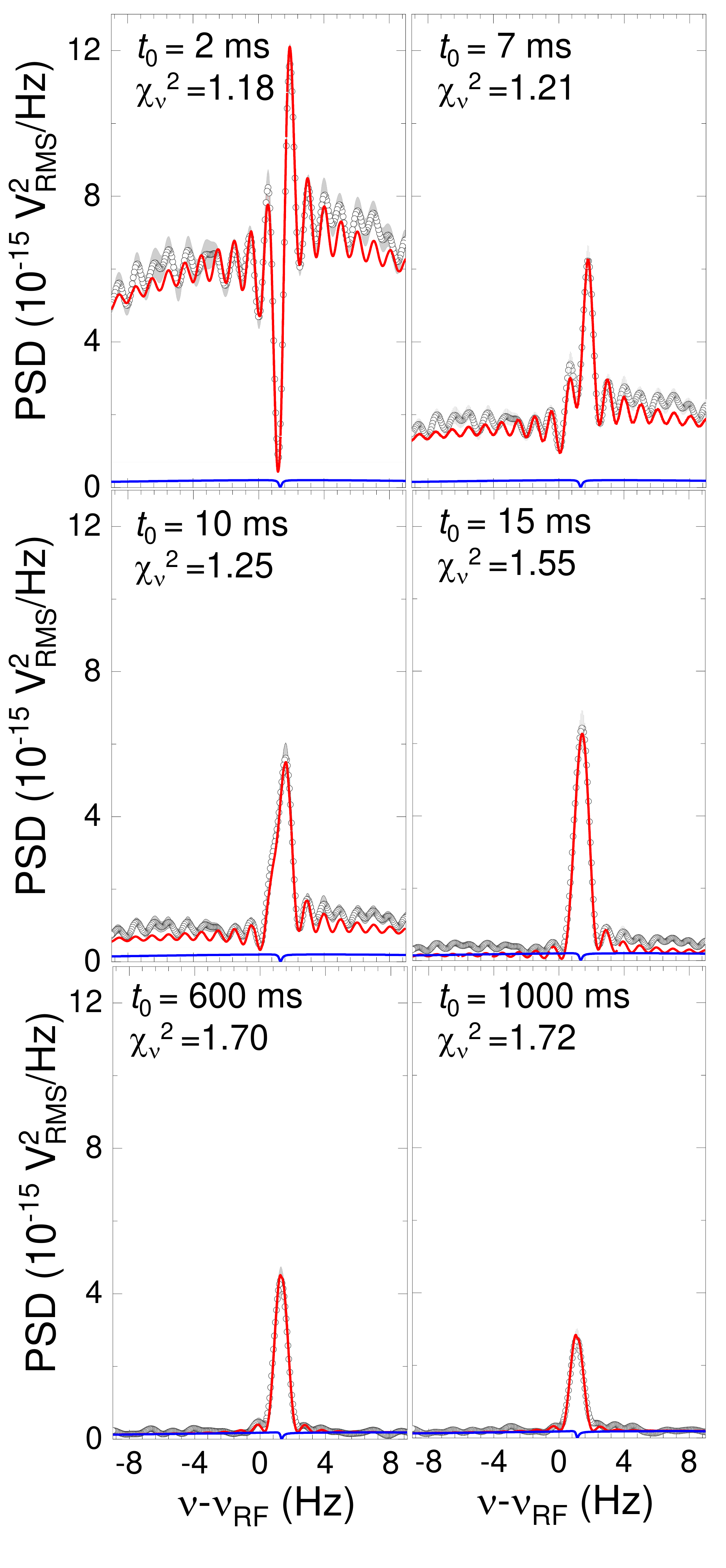}
\vspace{-7mm}
\caption{Evolution of the PSD signal and fitting function for different times $t_0$. The data points and standard deviations are the results from 20 measurements.}
\label{fig:evolution}
\end{figure}

The fit to the experimental data is done with $  S(\omega)$ (Eq.~(\ref{eq:fitting_function_general})). The terms, accounting for the coherent and thermal interaction, respectively, named as $S^{\text{coh}}(\omega)$ and $S^{\text{th}}(\omega)$, are displayed in Eqs.~(\ref{eq:coherent_power}) and (\ref{eq:thermal_power}). Many parameters have to be considered:
$S^{\text{noise}}$, $\gamma _{\hbox{\scriptsize{ion}}}$, $\gamma _{\hbox{\scriptsize{q}}}$, $\nu _{\hbox{\scriptsize{ion}}}$, $\nu _{\hbox{\scriptsize{q}}}$, $|g|$, $n _{\hbox{\scriptsize{ion}}}$, $n _{\hbox{\scriptsize{q}}}$, $|\langle a\rangle |$,  $|\langle b\rangle |$, $\theta _a$, $\theta _b$, $f_{\hbox{\scriptsize{ion}}}$, $f_{\hbox{\scriptsize{q}}}$, $\phi_{\hbox{\scriptsize{ion}}}$, and $\phi_{\hbox{\scriptsize{q}}}$. We consider $\theta_\text{b}=0$ because it can be factor out as a global phase and neglected due to the absolute value in $S^{\text{coh}}(\omega)$. Under this assumption, only the relative phase $\delta \approx \theta_\text{a}$ is relevant for the fit. To simplify the analysis, we only consider $t_0\geq 0$, to neglect the RF field related terms $f_{\hbox{\scriptsize{ion}}}$, $f_{\hbox{\scriptsize{q}}}$, $\phi_{\hbox{\scriptsize{ion}}}$, and $\phi_{\hbox{\scriptsize{q}}}$. 
A full fitting procedure consists of three steps described in the following:
\begin{itemize}
\item Fit of the background signal $S^{\text{th}}(\omega)+S^{\text{noise}}$ (without ions) using Eq.~(\ref{eq:thermal_power}) with $|g|=0$. This will yield the parameters $n _{\hbox{\scriptsize{q}}}$, $\gamma _{\hbox{\scriptsize{q}}}$ and $S^{\text{noise}}$. It will also yield $\nu _{\hbox{\scriptsize{q}}}$ although this will be considered as free parameter later. Figure~\ref{fig:fit_Nq} shows the evolution of $n_{\hbox{\scriptsize{q}}}$ for times $t_0$ in which the quartz is completely thermalized with its medium. The right panel shows the results for $\gamma _{\hbox{\scriptsize{q}}}$. $S^{\text{noise}}$ as a function of the time $t_0$ is shown in Appendix~\ref{other_quantities}.
\item Fit of the full signal $S(\omega)$ (with ions) for a fixed frequency value $\nu _1$, as a function of $t_0$, thus moving the integration window, to obtain \linebreak $|g|=2\pi \times 1.449(2)$~Hz and $\gamma _{\hbox{\scriptsize{ion}}}\approx 0$. $t_0=0$ is the time when the RF driving field is stopped. We consider $n _{\hbox{\scriptsize{ion}}}=n_{\hbox{\scriptsize{q}}}$ since in equilibrium, the resonator is at room temperature and its interaction with the ions is done through the detection electrode also at room temperature. Further details are given in Appendix~\ref{other_quantities}. 
\item Fit of the full signal $S(\omega)$ (with ions) considering as fixed parameters  $\gamma _{\hbox{\scriptsize{q}}}$, $\gamma _{\hbox{\scriptsize{ion}}}$, $n _{\hbox{\scriptsize{q}}}=n _{\hbox{\scriptsize{ion}}}$ and $\abs{g}$. Five parameters are obtained from the fit: $\delta$, $\nu _{\hbox{\scriptsize{ion}}}$, $\nu _{\hbox{\scriptsize{q}}}$, $\abs{\left\langle a(t_0) \right \rangle}$ and $\abs{\left \langle b(t_0) \right \rangle}$. 
\end{itemize}
The data points for about $N=6000$~ions \cite{Berr2021} and the fits for $t_0=0$ are shown in Fig.~\ref{fig:fit1}. Results considering other values of $t_0$ are presented in Fig.~\ref{fig:evolution}. The data acquisition window $t_d$ considered for the fits is always one~second, which is reasonable to obtain good resolution. For  $t_d \gtrsim 2$~s  the fit function $S(\omega)$ does not converge since $\nu_{\hbox{\scriptsize{ion}}}$ is not stable after some hundreds of miliseconds (see below). 
The effective relative phase (ERP) between quartz and trapped ions, which is defined as $\text{ERP} \equiv \pi/2 - \text{arg}\{g\}-\delta(t_0)$, because of the $ig^{*}\left\langle a^{\dagger}\left(t_{0}\right)\right\rangle =e^{i(\pi/2-\text{arg}\{g\}-\delta(t_{0}))}\left|g\right|\left|\left\langle a^{\dagger}\left(t_{0}\right)\right\rangle \right|\equiv e^{i\cdot\text{ERP}}\left|g\right|\left|\left\langle a^{\dagger}\left(t_{0}\right)\right\rangle \right|$ factor in Eq.~(\ref{eq:coherent_power}), is $\approx 150^{\hbox{\scriptsize{o}}}$ for $t_0$ varying from $0$ to $25$~ms. For $t_0 > 25$~ms, the ERP decreases gradually down to $0^{\hbox{\scriptsize{o}}}$ at $t_0=50$~ms. The different phase implies the subtraction or the sum of the Lorentzian functions quantified in Eq.~(\ref{eq:coherent_power}). This gives rise to a dip or a peak signal. The dip is only visible while the crystal is still energized after the driving.

\begin{figure}[t]
\includegraphics[scale=0.34]{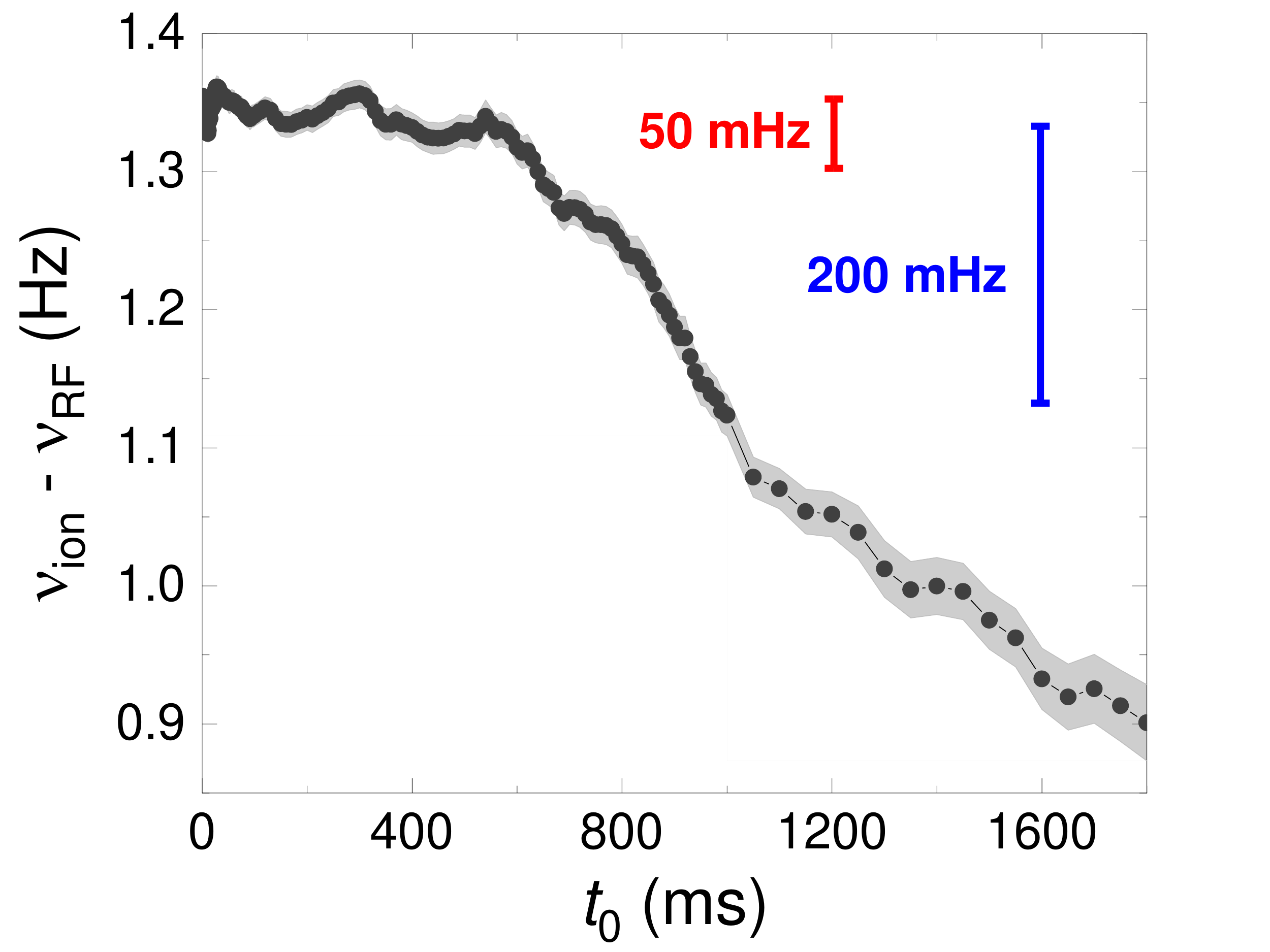}
\vspace{-7mm}
\caption{Evolution of $\nu _{\hbox{\scriptsize{ion}}}$ obtained from the fit using Eq.~(\ref{eq:fitting_function_general}) as a function of $t_0$.}
\label{fig:nu_+_results}
\end{figure}

The use of quartz crystals for high-precision Penning-trap mass spectrometry relies on how accurate motional frequencies can be determined. Figure~\ref{fig:nu_+_results} shows the evolution of $\nu _{\hbox{\scriptsize{ion}}}\approx \nu _+$ as a function of $t_0$. The analysis procedure developed in this manuscript results in an accurate value when $t_0$ is varied from 0 to 600~ms, improving the results based on Gaussian fits presented in Ref.~\cite{Berr2021}. The trend in the frequency towards lower values observed in Fig.~\ref{fig:nu_+_results} might be assigned to relaxation of the ion cloud ($\approx 6000$~ions in this experiment) after this has been driven. One would expect this trend to be less pronounced when reducing the ion number, and to be negligible in the limit of one detected ion.

\begin{figure}[b]
\includegraphics[scale=0.44]{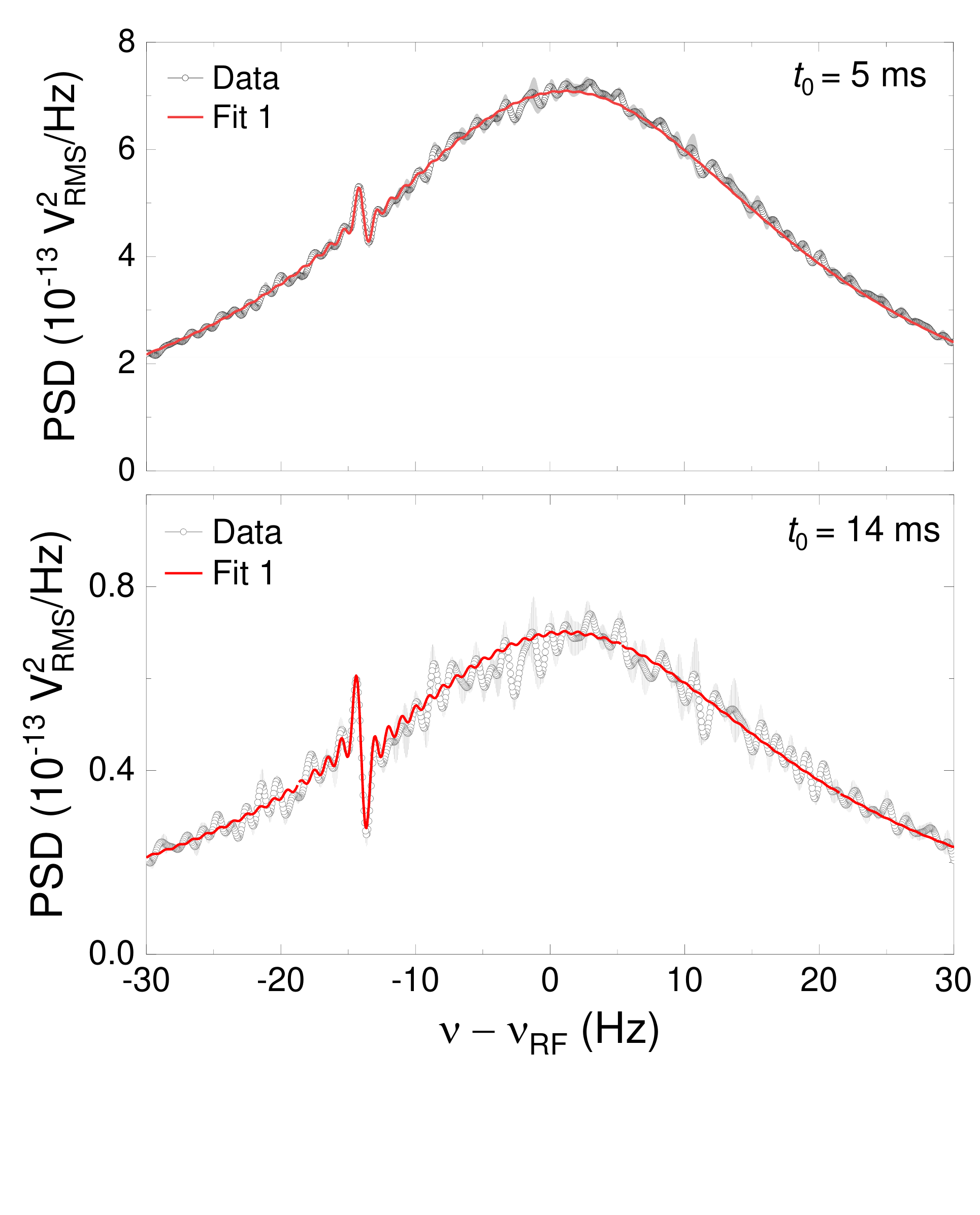}
\vspace{-21mm}
\caption{PSD signals as a function of the frequency for \linebreak {$t_0=5$~ms} (top) and $t_0=14$~ms (bottom). The red-solid line (Fit 1) is the $S(\omega)$ fit. The $S^{\text{th}}(\omega)$ fit is not shown since it is not visible in this scale. Note that this is two orders of magnitude compared to Fig.~\ref{fig:fit1} because the RF field amplitude applied here is 16 times larger. The data points and standard deviations are the results from 3 measurements.}
\label{fig:nu_+_results_lasers}
\end{figure}

Finally, the case where $\nu _{\hbox{\scriptsize{ion}}}$ is shifted from $\nu _{\hbox{\scriptsize{q}}}$ is outlined. Figure~\ref{fig:nu_+_results_lasers} shows the PSD signal as a function of the frequency for two different values of $t_0$. For these data the amplitude of the RF field applied was a factor of 16 larger, compared to that applied for the data shown in Figs.~\ref{fig:fit1} and \ref{fig:evolution}. This is because here, the ions were laser-cooled. A total of $750$~ions were trapped \cite{Berr2021}. For the presentation of these data, the analysis procedure has been shorten, considering $\gamma _{\hbox{\scriptsize{q}}}$ variable and using $|g|$ (scaled with the number of ions) obtained from the first set of data. Thus, the second step, which is presented in Appendix ~\ref{other_quantities} has been omitted. This yields fits with $\chi ^2$ very large, although for the purpose of this manuscript this is reasonable. 

\section{Conclusions and Outlook}

In this publication we have demonstrated that the hybrid system quartz-trapped-ions behaves as a coupled-oscillators system. The function to analyze the experimental power spectrum obtained for the charge induced by an ion cloud on trap electrodes has been fully developed. The evolution of the ions'  signal from a dip to a peak structure is explained due to a change in the relative phase between the two oscillators and the change in amplitude of the Lorentzian function used to describe the response of the quartz crystal. From the analysis procedure, we obtain the modified-cyclotron frequency, and the first motivation for our research, i.e., mass spectrometry has been discussed. Other important parameters can be also obtained, such as the ions-quartz coupling constant, determined to be as large as $|g|=2\pi \times 1.449(2)$~Hz in our experiments. From Eq.~(\ref{eq:acoplamiento_g}), it is possible to increase $g$ by reducing the distance between electrodes, by modifying the trap geometry quantified by $\alpha$, by increasing $N$ or $\omega _+$, or by reducing $C_{\hbox{\scriptsize{q}}}$. The latter would imply a larger quality factor for the quartz resonator. This quality factor varied in the experiments reported here from 66,800 to 60,700 depending on the energy stored in the crystal. Very recently, we have improved the quality factor of the system quartz+amplifier+trap, by a factor of $\approx 2$. Since the measured quality factor of the bare system quartz+amplifier is a factor of $\approx 500$ larger,  this subjects the possibility to increase the coupling constant, to improve the detection sensitivity and to broaden the applications of such system, as for example to delay the relaxation time of an ion cloud after this has been probed at any of its motional frequencies. Such a system can be used in setup involving ions stored in different traps following the original proposal by Heinzen and Wineland \cite{Hein1990}.

\section*{Acknowledgement}
We acknowledge support from the Spanish MCINN through the project PID2019-104093GB-I00/AEI/10.01339/501100011033 and contract PTA2018-016573-I, and from the Andalusian Government through the project P18-FR-3432 and Fondo Operativo FEDER A-FQM-425-UGR18, from the Spanish Ministry of Education through PhD fellowship FPU17/02596, and from the University of Granada "Plan propio - Programa de Intensificaci\'on de la Investigaci\'on", project PP2017-PRI.I-04 and "Laboratorios Singulares 2020". The construction of the facility was supported by the European Research Council (contract number 278648-TRAPSENSOR), projects FPA2015-67694-P and FPA2012-32076, infrastructure projects UNGR10-1E-501, UNGR13-1E-1830 and EQC2018-005130-P (MICINN/FEDER/UGR), and IE-5713 and IE2017-5513 (Junta de Andaluc\'ia-FEDER). 

\appendix

\section{Quantum description of the different subsystems} \label{quantum_description}

The quantum hamiltonians presented in Sec.~\ref{sec:model} are deduced from considerations given in this appendix.

\subsection{Quantum description of the ion motion in a Penning trap} \label{quantum_description_ion}

The quantum hamiltonian of an ion in a Penning trap is given by \cite{Crim2018}
\begin{widetext}
\begin{equation}
    H_{\hbox{\scriptsize{ion}}}=\frac{1}{2m_{\hbox{\scriptsize{ion}}}}\left(\hat{p}_{x}^{2}+\hat{p}_{y}^{2}+\hat{p}_{z}^{2}\right)+\frac{\omega_{c}}{2}\left(\hat{x}\hat{p}_{y}-\hat{y}\hat{p}_{x}\right)+\frac{1}{2}m_{\hbox{\scriptsize{ion}}}\frac{\omega_{c}^{2}-2\omega_{z}^{2}}{4}\left(\hat{x}^{2}+\hat{y}^{2}\right)+\frac{1}{2}m_{\hbox{\scriptsize{ion}}}\omega_{z}^{2}\hat{z}^{2}
\end{equation}
that can be rewritten in the form
\begin{equation}
    H_{\hbox{\scriptsize{ion}}}=H_{z}+H_{+}+H_{-}=\hbar\omega_{z}\left(a_{z}^{\dagger}a_{z}+\frac{1}{2}\right)+\hbar\omega_{+}\left(a_{+}^{\dagger}a_{+}+\frac{1}{2}\right)-\hbar\omega_{-}\left(a_{-}^{\dagger}a_{-}+\frac{1}{2}\right),
    \label{eqn:fund_Hion}
\end{equation}
after defining the Fock creation and annihilation operators $a^{\dagger}$ and $a$, respectively, for each of the eigenmotions of the ion in the trap; modified-cyclotron (+), magnetron (-) and axial (z). These operators read
\begin{equation}
    \begin{split}
         a_{z}=\sqrt{\frac{Nm_{\hbox{\scriptsize{ion}}}\omega_{z}}{2\hbar}}\left[\hat{z}+\frac{i}{m_{\hbox{\scriptsize{ion}}}\omega_{z}}\hat{p}_{z}\right]\,,\qquad & a_{\pm}=\sqrt{\frac{Nm_{\hbox{\scriptsize{ion}}}}{2\hbar\left(\omega_{+}-\omega_{-}\right)}}\left[\left(\frac{\hat{p}_{x}}{m_{\hbox{\scriptsize{ion}}}}-\omega_{\pm}\hat{y}\right)\pm i\left(\frac{\hat{p}_{y}}{m_{\hbox{\scriptsize{ion}}}}+\omega_{\pm}\hat{x}\right)\right],\\
    a_{z}^{\dagger}=\sqrt{\frac{Nm_{\hbox{\scriptsize{ion}}}\omega_{z}}{2\hbar}}\left[\hat{z}-\frac{i}{m_{\hbox{\scriptsize{ion}}}\omega_{z}}\hat{p}_{z}\right]\,,\qquad &a_{\pm}^{\dagger}=\sqrt{\frac{Nm_{\hbox{\scriptsize{ion}}}}{2\hbar\left(\omega_{+}-\omega_{-}\right)}}\left[\left(\frac{\hat{p}_{x}}{m_{\hbox{\scriptsize{ion}}}}-\omega_{\pm}\hat{y}\right)\mp i\left(\frac{\hat{p}_{y}}{m_{\hbox{\scriptsize{ion}}}}+\omega_{\pm}\hat{x}\right)\right].
    \end{split} \label{ecuaciones_a}
\end{equation}\end{widetext}
where we have replaced the single ion with mass $m_{\hbox{\scriptsize{ion}}}$, charge $q_{\hbox{\scriptsize{ion}}}$ and momentum $p$  by an ion cloud with $N$ ions, considering them as single ion with mass $Nm_{\hbox{\scriptsize{ion}}}$, charge $Nq_{\hbox{\scriptsize{ion}}}$ and momentum $Np$.

\subsection{Quantum description of a quartz crystal} \label{quantum_description_quartz}

In similar way as it has been done for a single ion, the operators to describe the characteristic voltage and intensity in the crystal can be written as
\begin{equation}
    \hat{V}\equiv\frac{\hat{Q}}{C_{\hbox{\scriptsize{q}}}}=\frac{1}{k}\sqrt{\frac{\hbar}{2m_{\hbox{\scriptsize{q}}}\omega_{\hbox{\scriptsize{q}}}}}\left(b^{\dagger}+b\right) =\frac{V_0}{2}\left(b^{\dagger}-b\right) \label{eq:V_b_1}
\end{equation}
and
\begin{equation}
\hat{I}=i\frac{C_{\hbox{\scriptsize{q}}}}{k}\sqrt{\frac{\hbar \omega_{\hbox{\scriptsize{q}}}}{2 m_{\hbox{\scriptsize{q}}}}}\left(b^{\dagger}-b\right).
\label{eq:I_b_1}
\end{equation}
The expected values can be expressed as
\begin{equation}
   \langle \hat{V}\rangle =\frac{1}{k}\sqrt{\frac{2\hbar}{m_{\hbox{\scriptsize{q}}}\omega_{\hbox{\scriptsize{q}}}}}\text{Re}\{\langle b \rangle \}\,,\qquad \langle \hat{I} \rangle =\frac{C_{\hbox{\scriptsize{q}}}}{k}\sqrt{\frac{\hbar \omega_{\hbox{\scriptsize{q}}}}{ m_{\hbox{\scriptsize{q}}}}}\text{Im}\{\langle b \rangle\}.
    \label{eq:V_b_2}
\end{equation}\\

\subsection{Quantum description of the system quartz-ions} \label{quantum_description_int}

In order to model the interaction between the ions and the quartz we assume for the latter a charge $\hat{Q}'=\hat{Q}+\hat{Q}_{\hbox{\scriptsize{ion}}}$ and a current $\hat{I}'=\hat{I}+\hat{I}_{\hbox{\scriptsize{ion}}}$, where $ \hat{Q}_{\hbox{\scriptsize{ion}}}$ and $I_{\hbox{\scriptsize{ion}}}$ have been defined in Eqs.~(\ref{eqn:x-axis}) and (\ref{eqn:induced_charge}), respectively.
$\hat{Q}$ e $\hat{I}$ are the charge and current associated to the quartz resonator, i.e., when there are no ions in the trap (or if they are not in resonance with the crystal). The Hamiltonians can be given as 
\begin{widetext}
\begin{equation}
    \begin{split}
        H_{\hbox{\scriptsize{q}}}'&=H_{I}'+H_{\hbox{\scriptsize{q}}}'=\left(\frac{k}{C_{\hbox{\scriptsize{q}}}}\right)^{2}\left[\frac{\left(\hat{I}+\hat{I}_{\hbox{\scriptsize{ion}}}\right)^{2}}{2m_{\hbox{\scriptsize{q}}}}+\frac{m_{\hbox{\scriptsize{q}}}\omega_{\hbox{\scriptsize{q}}}^{2}\left(\hat{Q}+\hat{Q}_{\hbox{\scriptsize{ion}}}\right)^{2}}{2}\right]=\ldots\\
        &=\left(\frac{k}{C_{\hbox{\scriptsize{q}}}}\right)^{2}\Bigg\{\Bigg. \frac{1}{2m_{\hbox{\scriptsize{q}}}}\left[\hat{I}^{2}+\left(\frac{Nq_{\hbox{\scriptsize{ion}}}\alpha}{2d_{0}m_{\hbox{\scriptsize{ion}}}}\right)^{2}\hat{p}_{x}^{2}+\left(-\frac{Nq_{\hbox{\scriptsize{ion}}}\alpha}{2d_{0}m_{\hbox{\scriptsize{ion}}}}\right)\hat{I}\hat{p}_{x}\right]\\
        &+\frac{m_{\hbox{\scriptsize{q}}}\omega_{\hbox{\scriptsize{q}}}^{2}}{2}\left[\hat{Q} ^{2}+\left(\frac{Nq_{\hbox{\scriptsize{ion}}}\alpha}{2d_{0}}\right)^{2}\hat{x}^{2}+\left(-\frac{Nq_{\hbox{\scriptsize{ion}}}\alpha}{2d_{0}}\right)\hat{Q }\hat{x}+\left(\frac{-q_{\hbox{\scriptsize{ion}}}\alpha}{2}\right)\hat{Q }+\left(\frac{q_{\hbox{\scriptsize{ion}}}^{2}\alpha^{2}}{2d_{0}}\right)\hat{x}+\left(\frac{q_{\hbox{\scriptsize{ion}}}\alpha}{2}\right)^{2}\right]\Bigg.\Bigg\} .
    \end{split}
\end{equation}
\end{widetext}
The terms accompanying $\hat{p}_{x}^{2}$ y $\hat{x}^{2}$ might modify slightly the oscillation frequency of the ions. This effect is considered negligible in the experiment. The new terms on $\hat{Q}$  and $\hat{x}$ will shift the zero potential of the two oscillators, also in a negligible amount. The term $\left(\frac{q_{\hbox{\scriptsize{ion}}}\alpha}{2}\right)^{2}$ is constant, and it will not appear in the equations of motion. 

\section{Proofs for the computation of the power spectrum}

\subsection{Same-time correlator equations}
\label{correlator}

The dynamics of the correlator matrix $\braket{A_m(t) A_n(t)}$ does not follow exactly the quantum regression theorem \cite{breuer}. The equation
\begin{align}
  &\mathrm{tr}\left\{A_mA_n\mathcal{L}_\text{tot}(O)\right\}
= -\frac{i}{\hbar}\mathrm{tr}\{[A_mA_n,H],O\}\\
  &\quad+ \frac{\gamma_q}{2}n_q
    \mathrm{tr}\{b[A_mA_n,b^\dagger]O + [b,A_mA_n]b^\dagger O\}
    \notag\\
  &\quad+ \frac{\gamma_q}{2}(n_q+1)
    \mathrm{tr}\{b^\dagger[A_mA_n,b]O + [b^\dagger, A_mA_n]bO\}.\notag
\end{align}
transforms into
\begin{align}
  \mathrm{tr}\left\{A_nA_m\mathcal{L}_\text{tot}(O)\right\}
  =& -M_{mr}\braket{A_rA_n} + F_m\braket{A_n} + \\
  &- M_{ns}\braket{A_mA_s} + \braket{A_m}F_n + C_{mn}\notag.
\end{align}
The real matrix $\mathbf{C}$ is defined as
\begin{align}
  C_{mn}
  &= \gamma_q (n_q[b,A_m][A_n,b^\dagger] + (n_q+1)[b^\dagger,A_m][A_n,b])\notag\\
  &+ \gamma_\text{ion}(n_\text{ion}[a,A_m][A_n,a^\dagger] +
(n_\text{ion}+1)[a^\dagger,A_m][A_n,a]),\notag
\end{align}
which evaluates to
\begin{align}
  \mathbf{C} &=
    \begin{pmatrix}
    0 & 0 & \gamma_\text{ion}(n_\text{ion} + 1) & 0 \\
    0 & 0 & 0 & \gamma_\text{q}(n_\text{q} + 1) \\
    \gamma_\text{ion}n_\text{ion} & 0 & 0 & 0 \\
    0 & \gamma_\text{q}n_\text{q} & 0 & 0
    \end{pmatrix}\notag\\
  &=
    \begin{pmatrix}
      0 & \pmb{\gamma} (\mathbf{n} + \openone) \\
      \pmb{\gamma} \mathbf{n} & 0
    \end{pmatrix}.
\end{align}

\subsection{Fourier-space propagator}
\label{sec:propagator}

The Fourier transform of the first order moments $\mathbf{\tilde{A}}$ are related to the initial values of the first order moments at $t_0$ via a propagator
\begin{equation}
  \mathbf{\tilde{A}}(\omega;t_1,t_0) = \frac{1}{\sqrt{t_1-t_0}} \mathbf{W}(-\omega,t_1-t_0)
  \braket{\mathbf{A}(t_0)}
\end{equation}
where we have introduced the Green function
\begin{align}
  \mathbf{W}(\pm\omega,t_d) & :=\int_{0}^{t_d} e^{-\mathbf{M}\tau}e^{\pm i \omega \tau}\dif{\tau}\\
  &= (\pm i\omega - \mathbf{M})^{-1}\left(e^{(\pm i\omega-\mathbf{M}) t_d}-\openone\right)\notag.
\end{align}
The Lorentzian prefactor to the exponential can be computed as the inverse of two $2\times2$ matrices
\begin{equation}
  (i\omega - \mathbf{M})^{-1} =
  \begin{pmatrix}
    (i\omega - \mathbf{m})^{-1} & 0 \\
    0 & (i \omega - \mathbf{m}^*)^{-1}
  \end{pmatrix},
\end{equation}
for positive and negative values of $\omega$. This equation can be written in terms of the Lorentzian envelope~\eqref{eq:envelope} and the matrix
\begin{equation}
  \mathbf{G}(\omega) =
  \begin{pmatrix}
     -\frac{\gamma_\text{q}}{2}+i(\omega-\tilde{\omega}_\text{q})
    & i g \\
    i g^*
    &  -\frac{\gamma_\text{ion}}{2}+i(\omega-\tilde{\omega}_\text{ion})
  \end{pmatrix},
\end{equation}
resulting
\begin{equation}
  \label{eq:inverse-M}
  (i\omega - \mathbf{M})^{-1} =
  \begin{pmatrix}
    F(\omega)\mathbf{G}(\omega) & 0 \\
    0 & F(-\omega)^* \mathbf{G}(-\omega)^*
  \end{pmatrix}.
\end{equation}


Note that $F(\omega)\simeq 0$ whenever $\omega < 0$, so that certain parts of the inverse $(\pm i \omega + \mathbf{M})^{-1}$ can be neglected depending on whether we choose the positive or negative sign accompanying $\omega$.

\subsection{Power spectrum simplifications}
\label{sec:power-spectrum-matrix}

It will be convenient to divide the power spectrum integral into two components, taking into account the ordering of time
\begin{align*}
  S^\text{th}_{mn}(\omega)
  &=\int_{t_0}^{t_1}\!\!\dif{\tau}_1\int_{t_0}^{\tau_1}\!\!\dif{\tau}_2
    e^{i\omega(\tau_1-\tau_2)} \braket{A_m(\tau_2)A_n(\tau_1)}\\
  &+\int_{t_0}^{t_1}\!\!\dif{\tau}_1\int_{\tau_1}^{t_1}\!\!\dif{\tau}_2
    e^{i\omega(\tau_1-\tau_2)} \braket{A_m(\tau_2)A_n(\tau_1)}.
\end{align*}
The integration limits can be reordered to bring out the lowest time
\begin{align*}
  S_{mn}(\omega)
  &=\frac{1}{t_d}\int_{t_0}^{t_1}\!\!\dif{\tau}_2\int_{\tau_2}^{t_1}\!\!\dif{\tau}_1
    e^{i\omega(\tau_1-\tau_2)}  \braket{A_m(\tau_2)A_n(\tau_1)}\\
  &+\frac{1}{t_d}\int_{t_0}^{t_1}\!\!\dif{\tau}_1\int_{\tau_1}^{t_1}\!\!\dif{\tau}_2
    e^{i\omega(\tau_1-\tau_2)} \braket{A_m(\tau_2)A_n(\tau_1)},
\end{align*}
and relabel adequately
\begin{align*}
  S_{mn}(\omega)
  &=\frac{1}{t_d}\int_{t_0}^{t_1}\!\!\dif{t}\int_{0}^{t_d-t}\!\!\dif{\tau} e^{i\omega \tau}
     \braket{A_m(t)A_n(t+s)}\\
  &+\frac{1}{t_d}\int_{t_0}^{t_1}\!\!\dif{t}\int_{0}^{t_d-t}\!\!\dif{\tau} e^{-i\omega\tau}
    \braket{A_m(t+s)A_n(t)}.
\end{align*}
The first integral can be manipulated, noting that
\begin{align}
  \braket{A_m(t) A_n(t+s)}
  &= \braket{A_n(t+s)^\dagger A_m(t)^\dagger}^*\\
  &= P_{nu}\braket{A_u(t+s) A_v(t)}^* P_{mv},
\end{align}
with some permutation matrices $\mathbf{P}$
\begin{equation}
  \mathbf{P} \mathbf{A} = \mathbf{A}^\dagger.
\end{equation}
By doing so, one arrives at a single computation
\begin{align*}
  \mathbf{S}(\omega)
  &= \mathbf{S}^+(\omega) + \mathbf{P} \mathbf{S}^+(\omega)^\dagger \mathbf{P},\mbox{ with}\\
 S^+_{mn}(\omega) &=\frac{1}{t_d}\int_{t_0}^{t_1}\!\!\dif{t}\int_{0}^{t_d-t}\!\!\dif{\tau} e^{-i\omega \tau}
     \braket{A_m(t+s)A_n(t)}.
\end{align*}
In the particular case of computing the autocorrelation function of the voltage, $\mathbf{P}\mathbf{v} = \mathbf{v},$ to have
\begin{equation}
  S_V(\omega) = 2 \mathrm{Re}\left( \mathbf{v}^{\hbox{\scriptsize{T}}} \mathbf{S}^+(\omega) \mathbf{v} \right).
\end{equation}

\subsection{Thermal state}  \label{thermal_state_appendix}

The thermal state's correlation matrix $T_{nm}=\braket{A_nA_m}_\text{th}$ reads
\begin{equation}
  \mathbf{T} =
  \begin{pmatrix}
    \braket{a^2} & \braket{ab} & \braket{a a^\dagger} & \braket{a b^\dagger}\\
    \braket{ba} & \braket{bb} & \braket{b a^\dagger} & \braket{b b^\dagger}\\
    \braket{a^\dagger a} & \braket{a^\dagger b} & \braket{a^\dagger a^\dagger} & \braket{a^\dagger b^\dagger}\\
    \braket{b^\dagger a} & \braket{b^\dagger b} & \braket{b^\dagger a^\dagger} & \braket{b^\dagger b^\dagger}
  \end{pmatrix}.
\end{equation}
It has a simple structure
\begin{equation}
  \label{eq:thermal}
  \mathbf{T} =
  \begin{pmatrix}
    0 & \mathbf{t}_{-\dagger} \\ \mathbf{t}_{\dagger-} & 0
  \end{pmatrix}, \mbox{ with }
  \mathbf{t}_{-\dagger} = \mathbf{t}_{\dagger-}+\openone,
\end{equation}
and satisfies the stationary equation
\begin{equation}
  \mathbf{M} \mathbf{T} + \mathbf{T} \mathbf{M}^{\hbox{\scriptsize{T}}} = \mathbf{C},
\end{equation}
which gives the two equivalent conditions
\begin{align}
  \mathbf{m} \mathbf{t}_{-\dagger} + \mathbf{t}_{-\dagger}\mathbf{m}^\dagger
  &= \pmb{\gamma}(\mathbf{n} + \openone) \\
  \mathbf{m}^* \mathbf{t}_{\dagger-} + \mathbf{t}_{\dagger-} \mathbf{m}^{\hbox{\scriptsize{T}}} &= \pmb{\gamma} \mathbf{n}.
\end{align}
The exact solution of these equations are given by
\begin{align}
  \braket{b^\dagger b}_\text{th} =
  &n_q + \frac{4 |g|^2(n_\text{ion} - n_\text{q}) \gamma_\text{ion} \Gamma_+}{4|g|^2\Gamma_+^2 + \gamma_\text{ion}\gamma_\text{q}(\Gamma_+^2+ 4\Omega_-^2)},\\
  \braket{a^\dagger b}_\text{th} =
  & - \frac{2 i g(n_\text{ion}-n_\text{q})(\Gamma_++2i\Omega_+)}{4|g|^2\Gamma_+^2+\gamma_\text{ion}\gamma_\text{q}(\Gamma_+^2+4\Omega_-^2)},\\
  \braket{a^\dagger a}_\text{th} =
  &n_\text{ion}-\frac{4|g|^2(n_\text{ion}-n_\text{q})\gamma_\text{q}\Gamma_+}{4|g|^2\Gamma_+^2+\gamma_\text{ion}\gamma_\text{q}(\Gamma_+^2+4\Omega_-^2)},
\end{align}
with the quantities $\Gamma_\pm  := \gamma_\text{ion}\pm \gamma_\text{q}$, and $\Omega_\pm := \omega_\text{ion}\pm \omega_\text{q}$.

\section{Other quantities from the fit} \label{other_quantities}

\begin{figure}[h]
\includegraphics[scale=0.34]{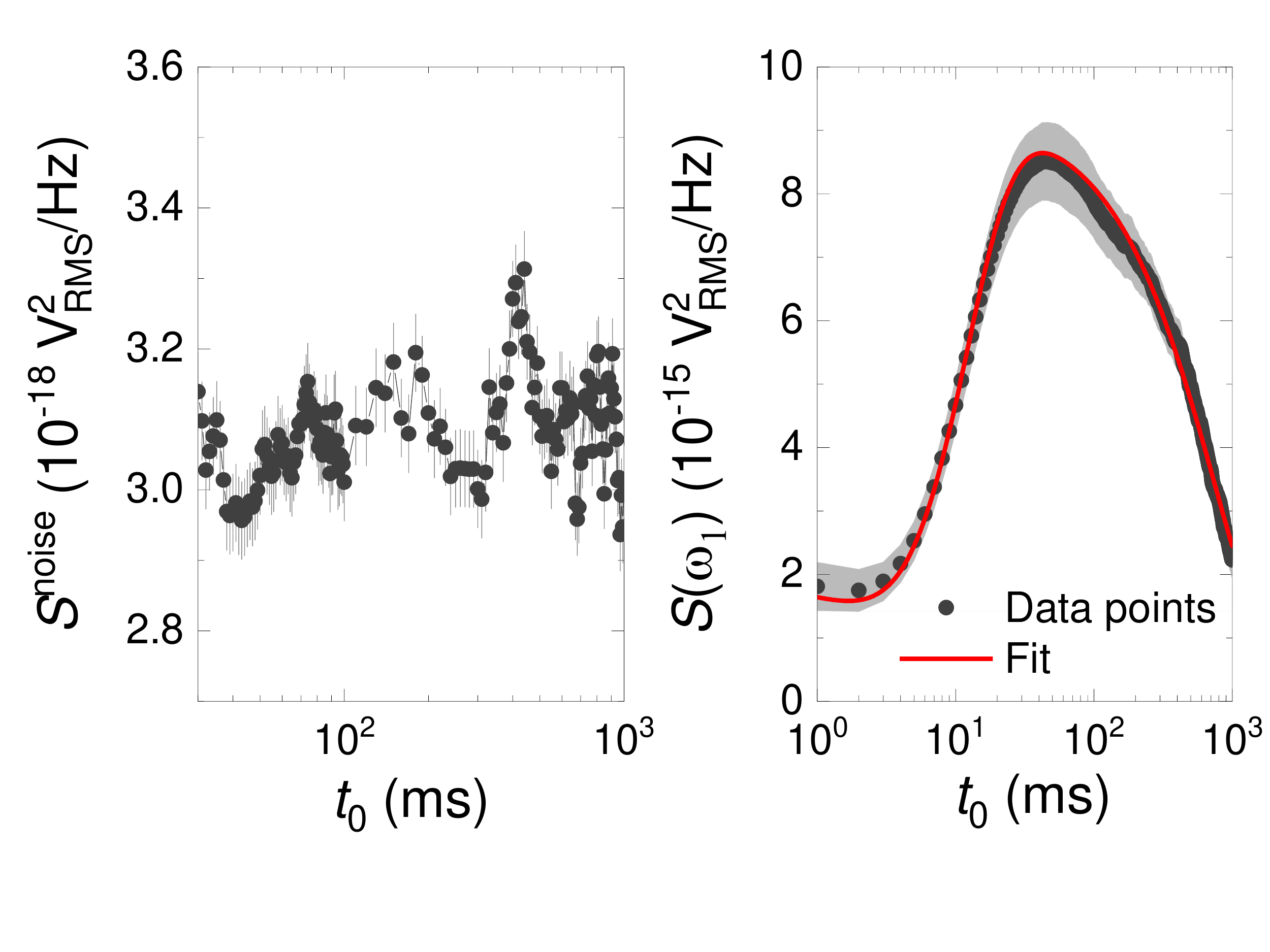}
\vspace{-14mm}
\caption{Left panel: $S^{\text{noise}}$. Right panel: $S(\omega) $ for a fixed frequency $\omega _1$.}
\label{fig:S_g}
\end{figure}

The part of the procedure that has been omitted in the main manuscript will be described in this section \cite{tfm_Emilio}. The left panel of Fig.~\ref{fig:S_g} shows the frequency-independent background contribution $S^{\text{noise}}$ (Eq.~(\ref{eq:fitting_function_general})).

\begin{figure}[h]
\includegraphics[scale=0.34]{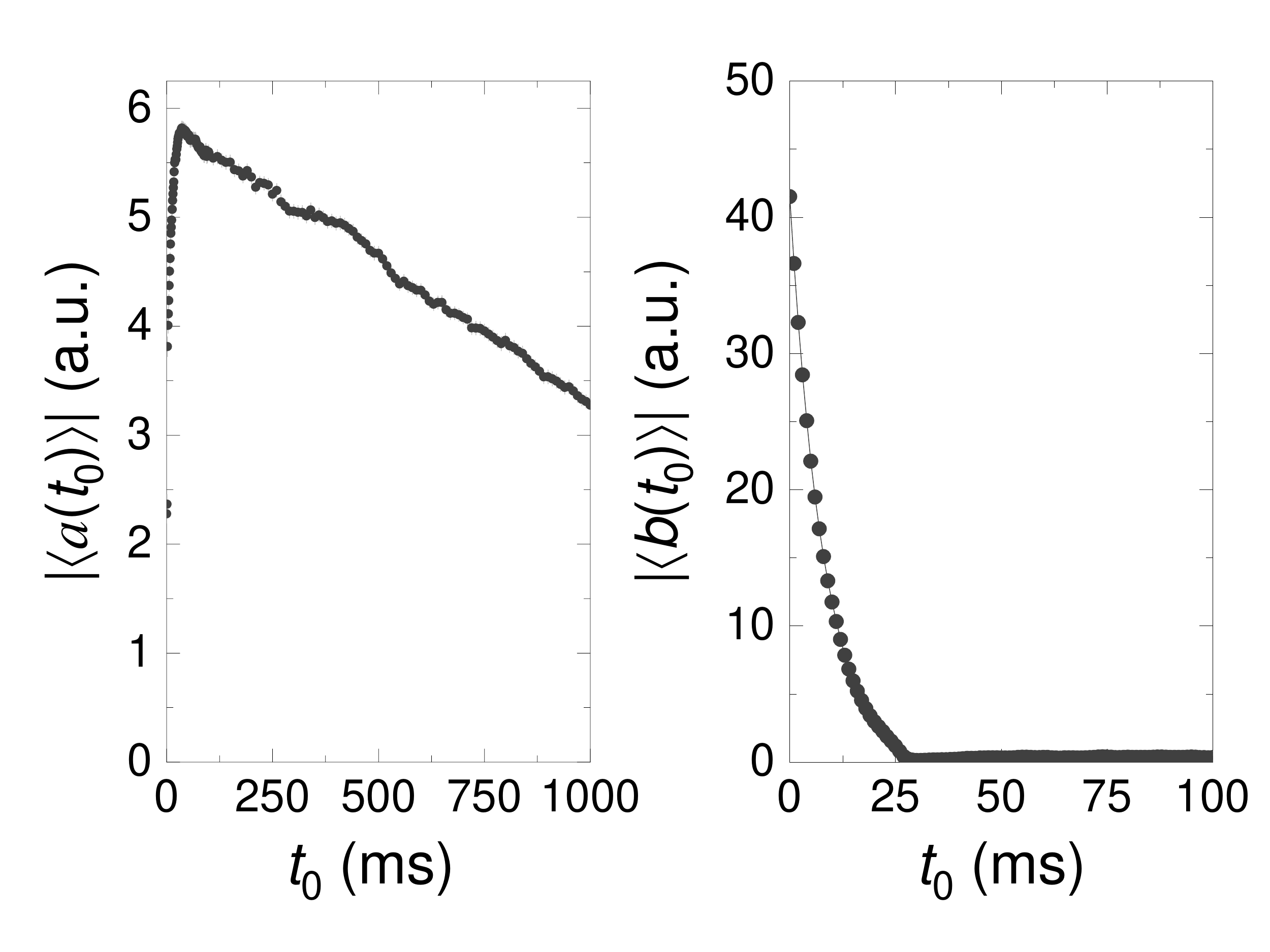}
\vspace{-10mm}
\caption{Time evolution of $|\langle a (t_0)\rangle |$ and $|\langle b(t_0) \rangle |$.}
\label{fig:a_b}
\end{figure}

In order to reduce the number of the free parameters gradually and considering the signal with and without ions the coupling constant $|g|$ has been obtained. The PSD near $\nu_+$ continuously decreases after \linebreak $t_0\gtrsim50$~ms as observed in the right panel of Fig.~\ref{fig:S_g}, when the quartz crystal amplitude $\left\langle b\left(t_{0}\right)\right\rangle$ is already $\sim 0$ (right panel of Fig.~\ref{fig:a_b}).
This happens because the ion cloud is losing energy after \linebreak{$t_0\gtrsim50$~ms} due to two main interactions:
\begin{itemize}
    \item With the quartz: ruled by their coupling constant $\abs{g}$, the quartz absorbs energy from the cloud and dissipates it in a (relatively short) time $\sim 1/\gamma_\text{q}$.
    \item With the ions' thermal bath: ruled by the dissipative constant $\gamma_\text{ion}$.
\end{itemize}

The parameters $\abs{g}$ and $\gamma_\text{ion}$ play a similar role in the ions' energy. It is hard to distinguish them from fits and get a stable parameters evolution. Therefore, $\gamma_\text{ion}$ is set to 0 in all fits to simplify the analysis. This implies that the value of $\abs{g}$ obtained in this paper should be considered as an upper limit.

Since $\abs{g}$ is considered one of the constant parameters in the model, the ideal computation would be to fit all signals simultaneously, for the different $t_0$,  setting $\abs{g}$ as a free parameter and searching for the minimum value of $\chi^2_\nu$. This has been discarded for simplicity. Instead, $\abs{g}$ has been obtained from a fit of the time evolution of the PSD at a single frequency value $\nu_1$ (right panel of Fig.~\ref{fig:S_g}). This has been arbitrarily chosen to be \linebreak $\nu_1=\nu_\text{RF}+1.35$ Hz because this value is almost at the peak maximum for signals with large $t_0$. The theoretical shape of this evolution is governed by  Eq.~(\ref{eq:fitting_function_general}) but fixing several parameters:

\begin{itemize}
    \item $\gamma_\text{ion}=0$, $\nu=\nu_1=\nu_\text{ion}=\nu_\text{RF}+1.35$ Hz.
    \item From the results of the background (no ions) fits: $S^{\text{noise}}=3.101(5)\times 10^{-18}~V_\text{RMS}^2$, $\nu_\text{q}(t_0=0\text{ ms})=\nu_\text{RF}+1.99(2)$ Hz, and \linebreak $\gamma_\text{q}(t_0=0\text{ ms})/2\pi=39.81(4)$ Hz.
    \item Solving Eq.~(\ref{eq:numerically}) numerically, we have obtained the values of $\left\langle a\left(t\right)\right\rangle $, $\left\langle a^\dagger\left(t\right)\right\rangle $, $\left\langle b\left(t\right)\right\rangle $ and $\left\langle b^\dagger\left(t\right)\right\rangle $ at \linebreak $t=t_1$ from its values at $t=t_0$. Figure~\ref{fig:a_b} shows the evolution of $|\langle a(t_0)\rangle|$ and $|\langle b(t_0)\rangle|$ as a function of $t_0$.
    \end{itemize}

Considering these values, Eq.~(\ref{eq:fitting_function_general}) is now a function of the independent variable $t_0$ and the free fit parameters $\abs{\left\langle a\left(t_{0}=0\right)\right\rangle}$, $\abs{\left\langle b\left(t_{0}=0\right)\right\rangle}$, $\abs{g}$, and the relative phase $\delta\left(t_{0}=0\right)$. This yields the upper limit \linebreak{$\abs{g}=2\pi \times 1.449(2)$~Hz}.

\begin{thebibliography}{24}%
\makeatletter
\providecommand \@ifxundefined [1]{%
 \@ifx{#1\undefined}
}%
\providecommand \@ifnum [1]{%
 \ifnum #1\expandafter \@firstoftwo
 \else \expandafter \@secondoftwo
 \fi
}%
\providecommand \@ifx [1]{%
 \ifx #1\expandafter \@firstoftwo
 \else \expandafter \@secondoftwo
 \fi
}%
\providecommand \natexlab [1]{#1}%
\providecommand \enquote  [1]{``#1''}%
\providecommand \bibnamefont  [1]{#1}%
\providecommand \bibfnamefont [1]{#1}%
\providecommand \citenamefont [1]{#1}%
\providecommand \href@noop [0]{\@secondoftwo}%
\providecommand \href [0]{\begingroup \@sanitize@url \@href}%
\providecommand \@href[1]{\@@startlink{#1}\@@href}%
\providecommand \@@href[1]{\endgroup#1\@@endlink}%
\providecommand \@sanitize@url [0]{\catcode `\\12\catcode `\$12\catcode
  `\&12\catcode `\#12\catcode `\^12\catcode `\_12\catcode `\%12\relax}%
\providecommand \@@startlink[1]{}%
\providecommand \@@endlink[0]{}%
\providecommand \url  [0]{\begingroup\@sanitize@url \@url }%
\providecommand \@url [1]{\endgroup\@href {#1}{\urlprefix }}%
\providecommand \urlprefix  [0]{URL }%
\providecommand \Eprint [0]{\href }%
\providecommand \doibase [0]{http://dx.doi.org/}%
\providecommand \selectlanguage [0]{\@gobble}%
\providecommand \bibinfo  [0]{\@secondoftwo}%
\providecommand \bibfield  [0]{\@secondoftwo}%
\providecommand \translation [1]{[#1]}%
\providecommand \BibitemOpen [0]{}%
\providecommand \bibitemStop [0]{}%
\providecommand \bibitemNoStop [0]{.\EOS\space}%
\providecommand \EOS [0]{\spacefactor3000\relax}%
\providecommand \BibitemShut  [1]{\csname bibitem#1\endcsname}%
\let\auto@bib@innerbib\@empty
\bibitem [{\citenamefont {Myers}(2019)}]{Myer2019}%
  \BibitemOpen
  \bibfield  {author} {\bibinfo {author} {\bibfnamefont {E.~G.}\ \bibnamefont
  {Myers}},\ }\href {\doibase 10.3390/atoms7010037} {\bibfield  {journal}
  {\bibinfo  {journal} {Atoms}\ }\textbf {\bibinfo {volume} {37}},\ \bibinfo
  {pages} {1} (\bibinfo {year} {2019})}\BibitemShut {NoStop}%
\bibitem [{\citenamefont {Borchert}\ \emph {et~al.}(2022)\citenamefont
  {Borchert}, \citenamefont {Devlin}, \citenamefont {Erlewein}, \citenamefont
  {Fleck}, \citenamefont {Harrington}, \citenamefont {Higuchi}, \citenamefont
  {Latacz}, \citenamefont {Voelksen}, \citenamefont {Wursten}, \citenamefont
  {Abbass}, \citenamefont {Bohman}, \citenamefont {Mooser}, \citenamefont
  {Popper}, \citenamefont {Wiesinger}, \citenamefont {Will}, \citenamefont
  {Blaum}, \citenamefont {Matsuda}, \citenamefont {Ospelkaus}, \citenamefont
  {Quint}, \citenamefont {Walz}, \citenamefont {Yamazaki}, \citenamefont
  {Smorra},\ and\ \citenamefont {Ulmer}}]{Borc2022}%
  \BibitemOpen
  \bibfield  {author} {\bibinfo {author} {\bibfnamefont {M.}~\bibnamefont
  {Borchert}}, \bibinfo {author} {\bibfnamefont {J.}~\bibnamefont {Devlin}},
  \bibinfo {author} {\bibfnamefont {S.}~\bibnamefont {Erlewein}}, \bibinfo
  {author} {\bibfnamefont {M.}~\bibnamefont {Fleck}}, \bibinfo {author}
  {\bibfnamefont {J.}~\bibnamefont {Harrington}}, \bibinfo {author}
  {\bibfnamefont {T.}~\bibnamefont {Higuchi}}, \bibinfo {author} {\bibfnamefont
  {B.}~\bibnamefont {Latacz}}, \bibinfo {author} {\bibfnamefont
  {F.}~\bibnamefont {Voelksen}}, \bibinfo {author} {\bibfnamefont
  {E.}~\bibnamefont {Wursten}}, \bibinfo {author} {\bibfnamefont
  {F.}~\bibnamefont {Abbass}}, \bibinfo {author} {\bibfnamefont
  {M.}~\bibnamefont {Bohman}}, \bibinfo {author} {\bibfnamefont
  {A.}~\bibnamefont {Mooser}}, \bibinfo {author} {\bibfnamefont
  {D.}~\bibnamefont {Popper}}, \bibinfo {author} {\bibfnamefont
  {M.}~\bibnamefont {Wiesinger}}, \bibinfo {author} {\bibfnamefont
  {C.}~\bibnamefont {Will}}, \bibinfo {author} {\bibfnamefont {K.}~\bibnamefont
  {Blaum}}, \bibinfo {author} {\bibfnamefont {Y.}~\bibnamefont {Matsuda}},
  \bibinfo {author} {\bibfnamefont {C.}~\bibnamefont {Ospelkaus}}, \bibinfo
  {author} {\bibfnamefont {W.}~\bibnamefont {Quint}}, \bibinfo {author}
  {\bibfnamefont {J.}~\bibnamefont {Walz}}, \bibinfo {author} {\bibfnamefont
  {Y.}~\bibnamefont {Yamazaki}}, \bibinfo {author} {\bibfnamefont
  {C.}~\bibnamefont {Smorra}}, \ and\ \bibinfo {author} {\bibfnamefont
  {S.}~\bibnamefont {Ulmer}},\ }\href {\doibase 10.1038/s41586-021-04203-w}
  {\bibfield  {journal} {\bibinfo  {journal} {Nature}\ }\textbf {\bibinfo
  {volume} {601}},\ \bibinfo {pages} {53} (\bibinfo {year} {2022})}\BibitemShut
  {NoStop}%
\bibitem [{\citenamefont {Hendrickson}\ \emph {et~al.}(2015)\citenamefont
  {Hendrickson}, \citenamefont {Quinn}, \citenamefont {Kaiser}, \citenamefont
  {Smith}, \citenamefont {Blakney}, \citenamefont {Chen}, \citenamefont
  {Marshall}, \citenamefont {Weisbrod},\ and\ \citenamefont {Beu}}]{Hend2015}%
  \BibitemOpen
  \bibfield  {author} {\bibinfo {author} {\bibfnamefont {C.~L.}\ \bibnamefont
  {Hendrickson}}, \bibinfo {author} {\bibfnamefont {J.~P.}\ \bibnamefont
  {Quinn}}, \bibinfo {author} {\bibfnamefont {N.~K.}\ \bibnamefont {Kaiser}},
  \bibinfo {author} {\bibfnamefont {D.~F.}\ \bibnamefont {Smith}}, \bibinfo
  {author} {\bibfnamefont {G.~T.}\ \bibnamefont {Blakney}}, \bibinfo {author}
  {\bibfnamefont {T.}~\bibnamefont {Chen}}, \bibinfo {author} {\bibfnamefont
  {A.~G.}\ \bibnamefont {Marshall}}, \bibinfo {author} {\bibfnamefont {C.~R.}\
  \bibnamefont {Weisbrod}}, \ and\ \bibinfo {author} {\bibfnamefont {S.~C.}\
  \bibnamefont {Beu}},\ }\href {\doibase 10.1007/s13361-015-1182-2} {\bibfield
  {journal} {\bibinfo  {journal} {J. Am. Soc. Mass Spectrom.}\ }\textbf
  {\bibinfo {volume} {1626-1632}},\ \bibinfo {pages} {26} (\bibinfo {year}
  {2015})}\BibitemShut {NoStop}%
\bibitem [{\citenamefont {Rischka}\ \emph {et~al.}(2020)\citenamefont
  {Rischka}, \citenamefont {Cakir}, \citenamefont {Door}, \citenamefont
  {Filianin}, \citenamefont {Harman}, \citenamefont {Huang}, \citenamefont
  {Indelicato}, \citenamefont {Keitel}, \citenamefont {K\"onig}, \citenamefont
  {Kromer}, \citenamefont {M\"uller}, \citenamefont {Novikov}, \citenamefont
  {Sch\"ussler}, \citenamefont {Schweiger}, \citenamefont {Eliseev},\ and\
  \citenamefont {Blaum}}]{Risc2020}%
  \BibitemOpen
  \bibfield  {author} {\bibinfo {author} {\bibfnamefont {A.}~\bibnamefont
  {Rischka}}, \bibinfo {author} {\bibfnamefont {H.}~\bibnamefont {Cakir}},
  \bibinfo {author} {\bibfnamefont {M.}~\bibnamefont {Door}}, \bibinfo {author}
  {\bibfnamefont {P.}~\bibnamefont {Filianin}}, \bibinfo {author}
  {\bibfnamefont {Z.}~\bibnamefont {Harman}}, \bibinfo {author} {\bibfnamefont
  {W.~J.}~\bibnamefont {Huang}}, \bibinfo {author} {\bibfnamefont
  {P.}~\bibnamefont {Indelicato}}, \bibinfo {author} {\bibfnamefont
  {C.~H.}~\bibnamefont {Keitel}}, \bibinfo {author} {\bibfnamefont
  {C.~M.}~\bibnamefont {K\"onig}}, \bibinfo {author} {\bibfnamefont
  {K.}~\bibnamefont {Kromer}}, \bibinfo {author} {\bibfnamefont
  {M.}~\bibnamefont {M\"uller}}, \bibinfo {author} {\bibfnamefont
  {Y.~N.}~\bibnamefont {Novikov}}, \bibinfo {author} {\bibfnamefont
  {R.~X.}~\bibnamefont {Sch\"ussler}}, \bibinfo {author} {\bibfnamefont
  {C.}~\bibnamefont {Schweiger}}, \bibinfo {author} {\bibfnamefont
  {S.}~\bibnamefont {Eliseev}}, \ and\ \bibinfo {author} {\bibfnamefont
  {K.}~\bibnamefont {Blaum}},\ }\href {\doibase 10.1103/PhysRevLett.124.113001}
  {\bibfield  {journal} {\bibinfo  {journal} {Phys. Rev. Lett.}\ }\textbf
  {\bibinfo {volume} {124}},\ \bibinfo {pages} {113001} (\bibinfo {year}
  {2020})}\BibitemShut {NoStop}%
\bibitem [{\citenamefont {Rainville}\ \emph {et~al.}(2004)\citenamefont
  {Rainville}, \citenamefont {Thompson},\ and\ \citenamefont
  {Pritchard}}]{Rain2004}%
  \BibitemOpen
  \bibfield  {author} {\bibinfo {author} {\bibfnamefont {S.}~\bibnamefont
  {Rainville}}, \bibinfo {author} {\bibfnamefont {J.~K.}\ \bibnamefont
  {Thompson}}, \ and\ \bibinfo {author} {\bibfnamefont {D.~E.}\ \bibnamefont
  {Pritchard}},\ }\href {\doibase 10.1126/science.109232} {\bibfield  {journal}
  {\bibinfo  {journal} {Science}\ }\textbf {\bibinfo {volume} {303}},\ \bibinfo
  {pages} {334} (\bibinfo {year} {2004})}\BibitemShut {NoStop}%
\bibitem [{\citenamefont {Weisskoff}\ \emph {et~al.}(1988)\citenamefont
  {Weisskoff}, \citenamefont {Lafyatis}, \citenamefont {Boyce}, \citenamefont
  {Cornell}, \citenamefont {Flanagan},\ and\ \citenamefont
  {Pritchard}}]{Weis1988}%
  \BibitemOpen
  \bibfield  {author} {\bibinfo {author} {\bibfnamefont {R.~M.}\ \bibnamefont
  {Weisskoff}}, \bibinfo {author} {\bibfnamefont {G.~P.}\ \bibnamefont
  {Lafyatis}}, \bibinfo {author} {\bibfnamefont {K.~R.}\ \bibnamefont {Boyce}},
  \bibinfo {author} {\bibfnamefont {E.~A.}\ \bibnamefont {Cornell}}, \bibinfo
  {author} {\bibfnamefont {R.~W.}\ \bibnamefont {Flanagan}}, \ and\ \bibinfo
  {author} {\bibfnamefont {D.}~\bibnamefont {Pritchard}},\ }\href@noop {}
  {\bibfield  {journal} {\bibinfo  {journal} {J.\ Appl.\ Phys.}\ }\textbf
  {\bibinfo {volume} {63}},\ \bibinfo {pages} {4599} (\bibinfo {year}
  {1988})}\BibitemShut {NoStop}%
\bibitem [{\citenamefont {Ulmer}\ \emph {et~al.}(2009)\citenamefont {Ulmer},
  \citenamefont {Kracke}, \citenamefont {Blaum}, \citenamefont {Kreim},
  \citenamefont {Mooser}, \citenamefont {Quint}, \citenamefont {Rodegheri},\
  and\ \citenamefont {Walz}}]{Ulme2009}%
  \BibitemOpen
  \bibfield  {author} {\bibinfo {author} {\bibfnamefont {S.}~\bibnamefont
  {Ulmer}}, \bibinfo {author} {\bibfnamefont {H.}~\bibnamefont {Kracke}},
  \bibinfo {author} {\bibfnamefont {K.}~\bibnamefont {Blaum}}, \bibinfo
  {author} {\bibfnamefont {S.}~\bibnamefont {Kreim}}, \bibinfo {author}
  {\bibfnamefont {A.}~\bibnamefont {Mooser}}, \bibinfo {author} {\bibfnamefont
  {W.}~\bibnamefont {Quint}}, \bibinfo {author} {\bibfnamefont {C.~C.}\
  \bibnamefont {Rodegheri}}, \ and\ \bibinfo {author} {\bibfnamefont
  {J.}~\bibnamefont {Walz}},\ }\href {\doibase 10.1063/1.3271537} {\bibfield
  {journal} {\bibinfo  {journal} {Rev.\ Sci.\ Instrum.}\ }\textbf {\bibinfo
  {volume} {80}},\ \bibinfo {pages} {123302} (\bibinfo {year}
  {2009})}\BibitemShut {NoStop}%
\bibitem [{\citenamefont {Lohse}\ \emph {et~al.}(2019)\citenamefont {Lohse},
  \citenamefont {Berrocal}, \citenamefont {Block}, \citenamefont {Chemarev},
  \citenamefont {Cornejo}, \citenamefont {Ram\'irez},\ and\ \citenamefont
  {Rodr\'iguez}}]{Lohs2019}%
  \BibitemOpen
  \bibfield  {author} {\bibinfo {author} {\bibfnamefont {S.}~\bibnamefont
  {Lohse}}, \bibinfo {author} {\bibfnamefont {J.}~\bibnamefont {Berrocal}},
  \bibinfo {author} {\bibfnamefont {M.}~\bibnamefont {Block}}, \bibinfo
  {author} {\bibfnamefont {S.}~\bibnamefont {Chemarev}}, \bibinfo {author}
  {\bibfnamefont {J.~M.}\ \bibnamefont {Cornejo}}, \bibinfo {author}
  {\bibfnamefont {J.~G.}\ \bibnamefont {Ram\'irez}}, \ and\ \bibinfo {author}
  {\bibfnamefont {D.}~\bibnamefont {Rodr\'iguez}},\ }\href {\doibase
  10.1063/1.5094428} {\bibfield  {journal} {\bibinfo  {journal} {Rev. Sci.
  Instrum.}\ }\textbf {\bibinfo {volume} {90}},\ \bibinfo {pages} {063202}
  (\bibinfo {year} {2019})}\BibitemShut {NoStop}%
\bibitem [{\citenamefont {Lohse}\ \emph {et~al.}(2020)\citenamefont {Lohse},
  \citenamefont {Berrocal}, \citenamefont {B\"ohland}, \citenamefont {van~de
  Laar}, \citenamefont {Block}, \citenamefont {Chemarev}, \citenamefont
  {D\"ullman}, \citenamefont {Nagy}, \citenamefont {Ram\'irez},\ and\
  \citenamefont {Rodr\'iguez}}]{Lohs2020}%
  \BibitemOpen
  \bibfield  {author} {\bibinfo {author} {\bibfnamefont {S.}~\bibnamefont
  {Lohse}}, \bibinfo {author} {\bibfnamefont {J.}~\bibnamefont {Berrocal}},
  \bibinfo {author} {\bibfnamefont {S.}~\bibnamefont {B\"ohland}}, \bibinfo
  {author} {\bibfnamefont {J.}~\bibnamefont {van~de Laar}}, \bibinfo {author}
  {\bibfnamefont {M.}~\bibnamefont {Block}}, \bibinfo {author} {\bibfnamefont
  {S.}~\bibnamefont {Chemarev}}, \bibinfo {author} {\bibfnamefont {C.~E.}\
  \bibnamefont {D\"ullman}}, \bibinfo {author} {\bibfnamefont {S.}~\bibnamefont
  {Nagy}}, \bibinfo {author} {\bibfnamefont {J.~G.}\ \bibnamefont {Ram\'irez}},
  \ and\ \bibinfo {author} {\bibfnamefont {D.}~\bibnamefont {Rodr\'iguez}},\
  }\href {\doibase 10.1063/5.0015011} {\bibfield  {journal} {\bibinfo
  {journal} {Rev. Sci. Instruments}\ }\textbf {\bibinfo {volume} {91}},\
  \bibinfo {pages} {093202} (\bibinfo {year} {2020})}\BibitemShut {NoStop}%
\bibitem [{\citenamefont {Berrocal}\ \emph {et~al.}(2021)\citenamefont
  {Berrocal}, \citenamefont {Lohse}, \citenamefont {Dom\'inguez}, \citenamefont
  {Guti\'errez}, \citenamefont {Fern\'andez}, \citenamefont {Block},
  \citenamefont {Garc\'ia-Ripoll},\ and\ \citenamefont
  {Rodr\'iguez}}]{Berr2021}%
  \BibitemOpen
  \bibfield  {author} {\bibinfo {author} {\bibfnamefont {J.}~\bibnamefont
  {Berrocal}}, \bibinfo {author} {\bibfnamefont {S.}~\bibnamefont {Lohse}},
  \bibinfo {author} {\bibfnamefont {F.}~\bibnamefont {Dom\'inguez}}, \bibinfo
  {author} {\bibfnamefont {M.~J.}\ \bibnamefont {Guti\'errez}}, \bibinfo
  {author} {\bibfnamefont {F.~J.}\ \bibnamefont {Fern\'andez}}, \bibinfo
  {author} {\bibfnamefont {M.}~\bibnamefont {Block}}, \bibinfo {author}
  {\bibfnamefont {J.~J.}\ \bibnamefont {Garc\'ia-Ripoll}}, \ and\ \bibinfo
  {author} {\bibfnamefont {D.}~\bibnamefont {Rodr\'iguez}},\ }\href {\doibase
  10.1088/2058-9565/ac01bc} {\bibfield  {journal} {\bibinfo  {journal} {Quantum
  Sci. Technol.}\ } (\bibinfo {year} {2021}),\
  10.1088/2058-9565/ac01bc}\BibitemShut {NoStop}%
\bibitem [{\citenamefont {Block}\ \emph {et~al.}(2010)\citenamefont {Block},
  \citenamefont {Ackermann}, \citenamefont {Blaum}, \citenamefont {Droese},
  \citenamefont {Dworschak}, \citenamefont {Eliseev}, \citenamefont
  {Fleckenstein}, \citenamefont {Haettner}, \citenamefont {Herfurth},
  \citenamefont {He{\ss}berger}, \citenamefont {Hofmann}, \citenamefont
  {Ketelaer}, \citenamefont {Ketter}, \citenamefont {Kluge}, \citenamefont
  {Marx}, \citenamefont {Mazzocco}, \citenamefont {Novikov}, \citenamefont
  {Pla{\ss}}, \citenamefont {Popeko}, \citenamefont {Rahaman}, \citenamefont
  {{Rodr{\'i}guez}}, \citenamefont {Scheidenberger}, \citenamefont
  {Schweikhard}, \citenamefont {Thirolf}, \citenamefont {Vorobyev},\ and\
  \citenamefont {Weber}}]{Bloc2010}%
  \BibitemOpen
  \bibfield  {author} {\bibinfo {author} {\bibfnamefont {M.}~\bibnamefont
  {Block}}, \bibinfo {author} {\bibfnamefont {D.}~\bibnamefont {Ackermann}},
  \bibinfo {author} {\bibfnamefont {K.}~\bibnamefont {Blaum}}, \bibinfo
  {author} {\bibfnamefont {C.}~\bibnamefont {Droese}}, \bibinfo {author}
  {\bibfnamefont {M.}~\bibnamefont {Dworschak}}, \bibinfo {author}
  {\bibfnamefont {S.}~\bibnamefont {Eliseev}}, \bibinfo {author} {\bibfnamefont
  {T.}~\bibnamefont {Fleckenstein}}, \bibinfo {author} {\bibfnamefont
  {E.}~\bibnamefont {Haettner}}, \bibinfo {author} {\bibfnamefont
  {F.}~\bibnamefont {Herfurth}}, \bibinfo {author} {\bibfnamefont {F.~P.}\
  \bibnamefont {He{\ss}berger}}, \bibinfo {author} {\bibfnamefont
  {S.}~\bibnamefont {Hofmann}}, \bibinfo {author} {\bibfnamefont
  {J.}~\bibnamefont {Ketelaer}}, \bibinfo {author} {\bibfnamefont
  {J.}~\bibnamefont {Ketter}}, \bibinfo {author} {\bibfnamefont {H.-J.}\
  \bibnamefont {Kluge}}, \bibinfo {author} {\bibfnamefont {G.}~\bibnamefont
  {Marx}}, \bibinfo {author} {\bibfnamefont {M.}~\bibnamefont {Mazzocco}},
  \bibinfo {author} {\bibfnamefont {Y.~N.}\ \bibnamefont {Novikov}}, \bibinfo
  {author} {\bibfnamefont {W.~R.}\ \bibnamefont {Pla{\ss}}}, \bibinfo {author}
  {\bibfnamefont {A.}~\bibnamefont {Popeko}}, \bibinfo {author} {\bibfnamefont
  {S.}~\bibnamefont {Rahaman}}, \bibinfo {author} {\bibfnamefont
  {D.}~\bibnamefont {{Rodr{\'i}guez}}}, \bibinfo {author} {\bibfnamefont
  {C.}~\bibnamefont {Scheidenberger}}, \bibinfo {author} {\bibfnamefont
  {L.}~\bibnamefont {Schweikhard}}, \bibinfo {author} {\bibfnamefont {P.~G.}\
  \bibnamefont {Thirolf}}, \bibinfo {author} {\bibfnamefont {G.~K.}\
  \bibnamefont {Vorobyev}}, \ and\ \bibinfo {author} {\bibfnamefont
  {C.}~\bibnamefont {Weber}},\ }\href {\doibase 10.1038/nature08774} {\bibfield
   {journal} {\bibinfo  {journal} {Nature}\ }\textbf {\bibinfo {volume}
  {463}},\ \bibinfo {pages} {785} (\bibinfo {year} {2010})}\BibitemShut
  {NoStop}%
\bibitem [{\citenamefont {Minaya~Ramirez}\ \emph {et~al.}(2012)\citenamefont
  {Minaya~Ramirez}, \citenamefont {Ackermann}, \citenamefont {Blaum},
  \citenamefont {Block}, \citenamefont {Droese}, \citenamefont
  {D{\"{u}}llmann}, \citenamefont {Dworschak}, \citenamefont {Eibach},
  \citenamefont {Eliseev}, \citenamefont {Haettner}, \citenamefont {Herfurth},
  \citenamefont {He{\ss}berger}, \citenamefont {Hofmann}, \citenamefont
  {Ketelaer}, \citenamefont {Marx}, \citenamefont {Mazzocco}, \citenamefont
  {Nesterenko}, \citenamefont {Novikov}, \citenamefont {Pla{\ss}},
  \citenamefont {Rodr{\'\i}guez}, \citenamefont {Scheidenberger}, \citenamefont
  {Schweikhard}, \citenamefont {Thirolf},\ and\ \citenamefont
  {Weber}}]{Mina2012}%
  \BibitemOpen
  \bibfield  {author} {\bibinfo {author} {\bibfnamefont {E.}~\bibnamefont
  {Minaya~Ramirez}}, \bibinfo {author} {\bibfnamefont {D.}~\bibnamefont
  {Ackermann}}, \bibinfo {author} {\bibfnamefont {K.}~\bibnamefont {Blaum}},
  \bibinfo {author} {\bibfnamefont {M.}~\bibnamefont {Block}}, \bibinfo
  {author} {\bibfnamefont {C.}~\bibnamefont {Droese}}, \bibinfo {author}
  {\bibfnamefont {{\relax Ch}.~E.}\ \bibnamefont {D{\"{u}}llmann}}, \bibinfo
  {author} {\bibfnamefont {M.}~\bibnamefont {Dworschak}}, \bibinfo {author}
  {\bibfnamefont {M.}~\bibnamefont {Eibach}}, \bibinfo {author} {\bibfnamefont
  {S.}~\bibnamefont {Eliseev}}, \bibinfo {author} {\bibfnamefont
  {E.}~\bibnamefont {Haettner}}, \bibinfo {author} {\bibfnamefont
  {F.}~\bibnamefont {Herfurth}}, \bibinfo {author} {\bibfnamefont {F.~P.}\
  \bibnamefont {He{\ss}berger}}, \bibinfo {author} {\bibfnamefont
  {S.}~\bibnamefont {Hofmann}}, \bibinfo {author} {\bibfnamefont
  {J.}~\bibnamefont {Ketelaer}}, \bibinfo {author} {\bibfnamefont
  {G.}~\bibnamefont {Marx}}, \bibinfo {author} {\bibfnamefont {M.}~\bibnamefont
  {Mazzocco}}, \bibinfo {author} {\bibfnamefont {D.}~\bibnamefont
  {Nesterenko}}, \bibinfo {author} {\bibfnamefont {{\relax Yu}.~N.}\
  \bibnamefont {Novikov}}, \bibinfo {author} {\bibfnamefont {W.~R.}\
  \bibnamefont {Pla{\ss}}}, \bibinfo {author} {\bibfnamefont {D.}~\bibnamefont
  {Rodr{\'\i}guez}}, \bibinfo {author} {\bibfnamefont {C.}~\bibnamefont
  {Scheidenberger}}, \bibinfo {author} {\bibfnamefont {L.}~\bibnamefont
  {Schweikhard}}, \bibinfo {author} {\bibfnamefont {P.~G.}\ \bibnamefont
  {Thirolf}}, \ and\ \bibinfo {author} {\bibfnamefont {C.}~\bibnamefont
  {Weber}},\ }\href {\doibase 10.1126/science.1225636} {\bibfield  {journal}
  {\bibinfo  {journal} {Science}\ }\textbf {\bibinfo {volume} {337}},\ \bibinfo
  {pages} {1207} (\bibinfo {year} {2012})}\BibitemShut {NoStop}%
\bibitem [{\citenamefont {Rodr\'iguez}\ \emph {et~al.}(2010)\citenamefont
  {Rodr\'iguez}, \citenamefont {Blaum}, \citenamefont {N\"ortersh\"auser},\
  and\ \citenamefont {\textit{et al.}}}]{Rodr2010}%
  \BibitemOpen
  \bibfield  {author} {\bibinfo {author} {\bibfnamefont {D.}~\bibnamefont
  {Rodr\'iguez}}, \bibinfo {author} {\bibfnamefont {K.}~\bibnamefont {Blaum}},
  \bibinfo {author} {\bibfnamefont {W.}~\bibnamefont {N\"ortersh\"auser}}, \
  and\ \bibinfo {author} {\bibnamefont {\textit{et al.}}},\ }\href {\doibase
  10.1140/epjst/e2010-01231-2} {\bibfield  {journal} {\bibinfo  {journal}
  {Eur.\ Phys.\ J.\ Special\ Topics}\ }\textbf {\bibinfo {volume} {183}},\
  \bibinfo {pages} {1} (\bibinfo {year} {2010})}\BibitemShut {NoStop}%
\bibitem [{\citenamefont {Wineland}\ and\ \citenamefont
  {Dehmelt}(1975)}]{Wine1975}%
  \BibitemOpen
  \bibfield  {author} {\bibinfo {author} {\bibfnamefont {D.~J.}\ \bibnamefont
  {Wineland}}\ and\ \bibinfo {author} {\bibfnamefont {H.~G.}\ \bibnamefont
  {Dehmelt}},\ }\href {\doibase 10.1063/1.321602} {\bibfield  {journal}
  {\bibinfo  {journal} {J.\ Appl.\ Phys.}\ }\textbf {\bibinfo {volume} {46}},\
  \bibinfo {pages} {919} (\bibinfo {year} {1975})}\BibitemShut {NoStop}%
\bibitem [{\citenamefont {Sturm}\ \emph {et~al.}(2017)\citenamefont {Sturm},
  \citenamefont {Arapoglou}, \citenamefont {Egl}, \citenamefont {H\"ocker},
  \citenamefont {Kraemer}, \citenamefont {Sailer}, \citenamefont {Tu},
  \citenamefont {Weigel}, \citenamefont {Wolf}, \citenamefont
  {L\'opez-Urrutia},\ and\ \citenamefont {Blaum}}]{Stur2019}%
  \BibitemOpen
  \bibfield  {author} {\bibinfo {author} {\bibfnamefont {S.}~\bibnamefont
  {Sturm}}, \bibinfo {author} {\bibfnamefont {I.}~\bibnamefont {Arapoglou}},
  \bibinfo {author} {\bibfnamefont {A.}~\bibnamefont {Egl}}, \bibinfo {author}
  {\bibfnamefont {M.}~\bibnamefont {H\"ocker}}, \bibinfo {author}
  {\bibfnamefont {S.}~\bibnamefont {Kraemer}}, \bibinfo {author} {\bibfnamefont
  {T.}~\bibnamefont {Sailer}}, \bibinfo {author} {\bibfnamefont
  {B.}~\bibnamefont {Tu}}, \bibinfo {author} {\bibfnamefont {A.}~\bibnamefont
  {Weigel}}, \bibinfo {author} {\bibfnamefont {R.}~\bibnamefont {Wolf}},
  \bibinfo {author} {\bibfnamefont {J.~C.}\ \bibnamefont {L\'opez-Urrutia}}, \
  and\ \bibinfo {author} {\bibfnamefont {K.}~\bibnamefont {Blaum}},\ }\href
  {\doibase 10.1140/epjst/e2018-800225-2} {\bibfield  {journal} {\bibinfo
  {journal} {Eur. Phys. J. ST}\ }\textbf {\bibinfo {volume} {227}},\ \bibinfo
  {pages} {1425} (\bibinfo {year} {2017})}\BibitemShut {NoStop}%
\bibitem [{\citenamefont {Rodr{\'i}guez}(2012)}]{Rodr2012}%
  \BibitemOpen
  \bibfield  {author} {\bibinfo {author} {\bibfnamefont {D.}~\bibnamefont
  {Rodr{\'i}guez}},\ }\href {\doibase 10.1007/s00340-011-4824-5} {\bibfield
  {journal} {\bibinfo  {journal} {Appl. Phys. B: Lasers O.}\ }\textbf {\bibinfo
  {volume} {107}},\ \bibinfo {pages} {1031} (\bibinfo {year}
  {2012})}\BibitemShut {NoStop}%
\bibitem [{\citenamefont {Bohman}\ \emph {et~al.}(2021)\citenamefont {Bohman},
  \citenamefont {Grunhofer}, \citenamefont {Smorra}, \citenamefont {Wiesinger},
  \citenamefont {Will}, \citenamefont {Borchert}, \citenamefont {Devlin},
  \citenamefont {Erlewein}, \citenamefont {Fleck}, \citenamefont {Gavranovic},
  \citenamefont {Harrington}, \citenamefont {Latacz}, \citenamefont {Mooser},
  \citenamefont {Popper}, \citenamefont {Wursten}, \citenamefont {Blaum},
  \citenamefont {Matsuda}, \citenamefont {Ospelkaus}, \citenamefont {Quint},
  \citenamefont {Walz}, \citenamefont {Ulmer},\ and\ \citenamefont
  {Collaboration}}]{Bohm2021}%
  \BibitemOpen
  \bibfield  {author} {\bibinfo {author} {\bibfnamefont {M.}~\bibnamefont
  {Bohman}}, \bibinfo {author} {\bibfnamefont {V.}~\bibnamefont {Grunhofer}},
  \bibinfo {author} {\bibfnamefont {C.}~\bibnamefont {Smorra}}, \bibinfo
  {author} {\bibfnamefont {M.}~\bibnamefont {Wiesinger}}, \bibinfo {author}
  {\bibfnamefont {C.}~\bibnamefont {Will}}, \bibinfo {author} {\bibfnamefont
  {M.~J.}\ \bibnamefont {Borchert}}, \bibinfo {author} {\bibfnamefont {J.~A.}\
  \bibnamefont {Devlin}}, \bibinfo {author} {\bibfnamefont {S.}~\bibnamefont
  {Erlewein}}, \bibinfo {author} {\bibfnamefont {M.}~\bibnamefont {Fleck}},
  \bibinfo {author} {\bibfnamefont {S.}~\bibnamefont {Gavranovic}}, \bibinfo
  {author} {\bibfnamefont {J.}~\bibnamefont {Harrington}}, \bibinfo {author}
  {\bibfnamefont {B.}~\bibnamefont {Latacz}}, \bibinfo {author} {\bibfnamefont
  {A.}~\bibnamefont {Mooser}}, \bibinfo {author} {\bibfnamefont
  {D.}~\bibnamefont {Popper}}, \bibinfo {author} {\bibfnamefont
  {E.}~\bibnamefont {Wursten}}, \bibinfo {author} {\bibfnamefont
  {K.}~\bibnamefont {Blaum}}, \bibinfo {author} {\bibfnamefont
  {Y.}~\bibnamefont {Matsuda}}, \bibinfo {author} {\bibfnamefont
  {C.}~\bibnamefont {Ospelkaus}}, \bibinfo {author} {\bibfnamefont
  {W.}~\bibnamefont {Quint}}, \bibinfo {author} {\bibfnamefont
  {J.}~\bibnamefont {Walz}}, \bibinfo {author} {\bibfnamefont {S.}~\bibnamefont
  {Ulmer}}, \ and\ \bibinfo {author} {\bibfnamefont {B.}~\bibnamefont
  {Collaboration}},\ }\href {\doibase 10.1038/s41586-021-03784-w} {\bibfield
  {journal} {\bibinfo  {journal} {Nature}\ }\textbf {\bibinfo {volume} {596}},\
  \bibinfo {pages} {514} (\bibinfo {year} {2021})}\BibitemShut {NoStop}%
\bibitem [{\citenamefont {Kotler}\ \emph {et~al.}(2017)\citenamefont {Kotler},
  \citenamefont {Simmonds}, \citenamefont {Leibfried},\ and\ \citenamefont
  {Wineland}}]{Kotl2017}%
  \BibitemOpen
  \bibfield  {author} {\bibinfo {author} {\bibfnamefont {S.}~\bibnamefont
  {Kotler}}, \bibinfo {author} {\bibfnamefont {R.~W.}\ \bibnamefont
  {Simmonds}}, \bibinfo {author} {\bibfnamefont {D.}~\bibnamefont {Leibfried}},
  \ and\ \bibinfo {author} {\bibfnamefont {D.~J.}\ \bibnamefont {Wineland}},\
  }\href {\doibase 10.1103/PhysRevA.95.022327} {\bibfield  {journal} {\bibinfo
  {journal} {Phys. Rev. A}\ }\textbf {\bibinfo {volume} {95}},\ \bibinfo
  {pages} {022327} (\bibinfo {year} {2017})}\BibitemShut {NoStop}%
\bibitem [{\citenamefont {Brown}\ and\ \citenamefont
  {Gabrielse}(1986)}]{Brow1986}%
  \BibitemOpen
  \bibfield  {author} {\bibinfo {author} {\bibfnamefont {L.~S.}\ \bibnamefont
  {Brown}}\ and\ \bibinfo {author} {\bibfnamefont {G.}~\bibnamefont
  {Gabrielse}},\ }\href {\doibase 10.1103/RevModPhys.58.233} {\bibfield
  {journal} {\bibinfo  {journal} {Rev. Mod. Phys.}\ }\textbf {\bibinfo {volume}
  {58}},\ \bibinfo {pages} {233} (\bibinfo {year} {1986})}\BibitemShut
  {NoStop}%
\bibitem [{\citenamefont {Clerk}\ \emph {et~al.}(2010)\citenamefont {Clerk},
  \citenamefont {Devoret}, \citenamefont {Girvin}, \citenamefont {Marquardt},\
  and\ \citenamefont {Schoelkopf}}]{quantum_noise_introduction}%
  \BibitemOpen
  \bibfield  {author} {\bibinfo {author} {\bibfnamefont {A.~A.}\ \bibnamefont
  {Clerk}}, \bibinfo {author} {\bibfnamefont {M.~H.}\ \bibnamefont {Devoret}},
  \bibinfo {author} {\bibfnamefont {S.~M.}\ \bibnamefont {Girvin}}, \bibinfo
  {author} {\bibfnamefont {F.}~\bibnamefont {Marquardt}}, \ and\ \bibinfo
  {author} {\bibfnamefont {R.~J.}\ \bibnamefont {Schoelkopf}},\ }\href
  {\doibase 10.1103/RevModPhys.82.1155} {\bibfield  {journal} {\bibinfo
  {journal} {Rev. Mod. Phys.}\ }\textbf {\bibinfo {volume} {82}},\ \bibinfo
  {pages} {1155} (\bibinfo {year} {2010})}\BibitemShut {NoStop}%
\bibitem [{\citenamefont {Breuer}\ and\ \citenamefont
  {Petruccione}(2010)}]{breuer}%
  \BibitemOpen
  \bibfield  {author} {\bibinfo {author} {\bibfnamefont {H.-P.}\ \bibnamefont
  {Breuer}}\ and\ \bibinfo {author} {\bibfnamefont {F.}~\bibnamefont
  {Petruccione}},\ }\href@noop {} {\emph {\bibinfo {title} {The theory of Open
  Quantum Systems}}}\ (\bibinfo  {publisher} {Oxford University Press},\
  \bibinfo {year} {2010})\BibitemShut {NoStop}%
\bibitem [{\citenamefont {Heinzen}\ and\ \citenamefont
  {Wineland}(1990)}]{Hein1990}%
  \BibitemOpen
  \bibfield  {author} {\bibinfo {author} {\bibfnamefont {D.~J.}\ \bibnamefont
  {Heinzen}}\ and\ \bibinfo {author} {\bibfnamefont {D.~J.}\ \bibnamefont
  {Wineland}},\ }\href {\doibase 10.1103/PhysRevA.42.2977} {\bibfield
  {journal} {\bibinfo  {journal} {Phys. Rev. A}\ }\textbf {\bibinfo {volume}
  {42}},\ \bibinfo {pages} {2977} (\bibinfo {year} {1990})}\BibitemShut
  {NoStop}%
\bibitem [{\citenamefont {Crimin}\ \emph {et~al.}(2018)\citenamefont {Crimin},
  \citenamefont {Garraway},\ and\ \citenamefont {Verd\'u.}}]{Crim2018}%
  \BibitemOpen
  \bibfield  {author} {\bibinfo {author} {\bibfnamefont {F.}~\bibnamefont
  {Crimin}}, \bibinfo {author} {\bibfnamefont {B.}~\bibnamefont {Garraway}}, \
  and\ \bibinfo {author} {\bibfnamefont {J.}~\bibnamefont {Verd\'u.}},\ }\href
  {\doibase 10.1080/09500340.2017.1393570} {\bibfield  {journal} {\bibinfo
  {journal} {J. Mod. Optics}\ }\textbf {\bibinfo {volume} {65}},\ \bibinfo
  {pages} {427} (\bibinfo {year} {2018})}\BibitemShut {NoStop}%
\bibitem [{\citenamefont {{E. Altozano}}(2021)}]{tfm_Emilio}%
  \BibitemOpen
  \bibfield  {author} {\bibinfo {author} {\bibnamefont {{E. Altozano}}},\
  }\emph {\bibinfo {title} {Estudio de resonadores de cuarzo para experimentos
  con trampas Penning}},\ \href@noop {} {Master's thesis},\ \bibinfo  {school}
  {Universidad de Granada} (\bibinfo {year} {2021})\BibitemShut {NoStop}%
\end{thebibliography}
\providecommand{\noopsort}[1]{}\providecommand{\singleletter}[1]{#1}%
\end{document}